\documentclass[pra,twocolumn,superscriptaddress,amsmath,amssymb,floatfix,noshowpacs]{revtex4}
\usepackage{graphicx,hyperref}
\usepackage{mathtools}
\usepackage{amsmath}
\usepackage{empheq}
\usepackage{tabularx}
\graphicspath{{fig/}}

\begin{document}
\title{\textbf{Theory of cavity QED with 2D atomic arrays}}
%\date{\today}
\author{Ephraim Shahmoon}
\affiliation{Department of Chemical \& Biological Physics, Weizmann Institute of Science, Rehovot 761001, Israel}
\affiliation{Department of Physics, Harvard University, Cambridge, Massachusetts 02138, USA}
\author{Dominik S.~Wild}
\affiliation{Department of Physics, Harvard University, Cambridge, Massachusetts 02138, USA}
\author{Mikhail D.~Lukin}
\affiliation{Department of Physics, Harvard University, Cambridge, Massachusetts 02138, USA}
\author{Susanne F.~Yelin}
\affiliation{Department of Physics, Harvard University, Cambridge, Massachusetts 02138, USA}
\affiliation{Department of Physics, University of Connecticut, Storrs, Connecticut 06269, USA}
\date{\today}

\begin{abstract}
We develop a quantum optical formalism to treat a two-dimensional array of atoms placed in an optical cavity. Importantly, and in contrast to typical treatments, we account for cooperative dipole-dipole effects mediated by the interaction of the atoms with the outside, non-cavity-confined modes. Based on the observation that scattering to these modes is largely inhibited due to these cooperative effects, we construct a generic formalism, independent of the specific cavity structure, and apply it to an array of non-saturated atoms. By further considering the atomic motion, we show that the inhibited damping can lead to a favorable scaling of the optomechanical parameters of an atom-array membrane placed within a cavity. The developed formalism lays the basis for further investigation of many-body QED with atom arrays in transversely confined geometries.
\end{abstract}

 \maketitle

\begin{figure}[t]
  \begin{center}
    \includegraphics[width=\columnwidth]{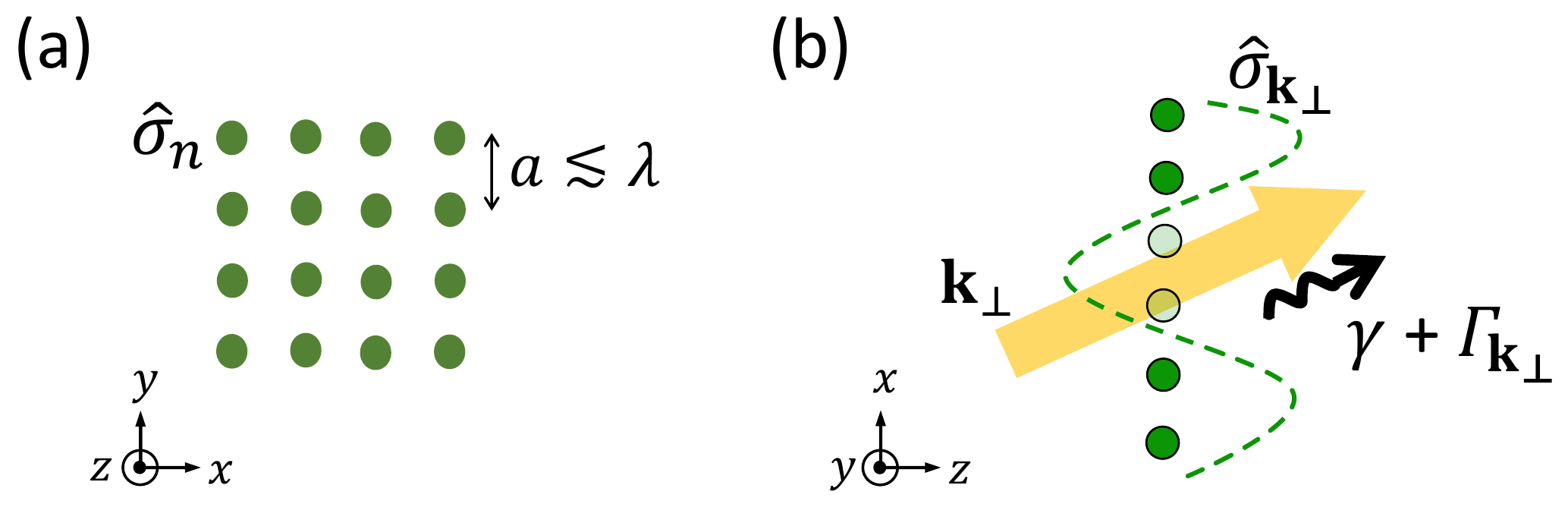}
    \caption{\small{
Light-matter interaction of a 2D atom array. (a) An array of two-level atoms ($\hat{\sigma}_n$ for atom $n$) spans the $xy$ plane with lattice spacing $a\lesssim\lambda$, $\lambda$ being the wavelength of the relevant light mode. (b) Dipole-dipole interactions between the array atoms lead to the formation of collective dipole modes, $\hat{\sigma}_{\mathbf{k}_{\bot}}$, characterized by the wavevector $\mathbf{k}_{\bot}=(k_x,k_y)$ [Eq. (\ref{sigk})]. Correspondingly, radiation is directed from these collective dipoles only to field modes with a matching wavevector $\mathbf{k}_{\bot}$ at a cooperative emission rate $\gamma+\Gamma_{\mathbf{k}_{\bot}}$.
}} \label{fig1}
  \end{center}
\end{figure}

\section{Introduction}
Quantum optical platforms which exhibit strong light-matter interactions play a crucial role in the study and application of various quantum phenomena and technologies. Spatially ordered arrays of laser-trapped atoms, as can be realized e.g. in optical lattices or tweezer arrays \cite{OL,QGM,BRW1,LUK,BRW2}, were recently considered as promising such platforms on the basis of their cooperative response to light. For lattice constant smaller than the wavelength of light (Fig. 1a), dipole-dipole interactions between the atoms become important. This, together with the lattice translational symmetry, gives rise to collective dipole excitations of the array whose interaction with light is highly directional (Fig. 1b). Strong light-matter interaction then occurs for light frequencies close to the cooperative resonance associated with the collective dipole modes, as can be characterized by the extremely strong reflectivity of a two-dimensional (2D) array, which was recently predicted \cite{ADM,coop} and also observed experimentally in an optical lattice \cite{BLO}. Other interesting potential applications include efficient couplers of collimated photons to single atoms \cite{coop,GRA}, enhancement of quantum memories and clocks \cite{ANA,MAN,HEN}, quantum communication \cite{GRA,ZOL}, waveguiding and subradiance \cite{coop,ANA,RITz,ANAc}, topological photonics \cite{janos,ADM2,janos3}, lasing \cite{ABA}, and optomechanics \cite{om,AAMO}

A different, well-established technique for achieving strong light-matter interaction is that of cavity quantum electrodynamics (QED), where atoms are trapped inside or near an optical cavity. For an ensemble of many atoms strongly interacting with a common, cavity-confined mode, this has led to the exploration and application of strong collective effects in e.g. many-body physics \cite{RITr,ESS1,ESS2,RIT1,RIT2,SAR}, quantum optomechanics \cite{SK1}, and quantum metrology \cite{VULh2,TOMh}. From the theory side, in contrast to the collective treatment of the interaction of atoms with the common \emph{confined} cavity mode, their interaction with the outside, \emph{non-cavity-confined} modes is typically considered at the individual atom level, resulting in a competing incoherent process of dipole-like emission and damping from each atom (Fig. 2a). However, when the atoms are close enough to each other, multiple scattering (via dipole-dipole interactions) prevails, and collective effects due to the interaction even with the outside modes becomes important. This may have dramatic effects especially when the atoms are spatially ordered, where the emission can become highly directional (Fig. 1b), so that the damping via scattering to outside, non-confined modes may be completely eliminated (Fig. 2b). The latter suggests that many-atom cavity QED using ordered atomic arrays may lead to new regimes and opportunities.

The goal of this paper is to provide a general theoretical framework for the description of cavity QED with ordered 2D atomic arrays. Such a formalism has to account for collectivity of atom-photon interactions at the level of both the confined cavity modes and the outside, non-confined modes. The treatment of the latter, collective interaction with non-confined modes, in an approximate manner which is independent of the cavity structure, is the main challenge addressed by this work. Restricting the current work to the case of non-saturated atoms, we illustrate our formalism by analyzing the optomechanical response of the system. We find favorable scalings of the corresponding optomechanical parameters compared to the case of disordered atomic ensembles. The latter is a direct consequence of the directionality of the cooperative emission, demonstrating the significance of the developed approach.

The paper is organized as follows. After introducing the model (Sec. II) and the resulting dynamical equations for a motionless ordered array of non-saturated atoms (Sec. III), we move to the heart of our formalism in Sec. IV. The latter contains a discussion of the generic physical approximation that allows for a consistent treatment of cooperative effects in atom arrays due to the non-confined modes. Application of this approximation to the array-cavity dynamics for motionless atoms is briefly discussed.
The rest of the paper is devoted to illustrating our approach when atomic motion is included, leading to an analysis of the optomechanics of the system. After revisiting the dynamical equations including motion (Sec. V), we focus on the case of large atom-cavity detuning (Sec. VI). This leads to a multimode optomechanical description of the system (Sec. VII), which is subsequently mapped onto the standard cavity optomechanics model (Sec. VIII). Finally, our conclusions are presented in Sec. IX.

\begin{figure}[t]
  \begin{center}
    \includegraphics[scale=0.45]{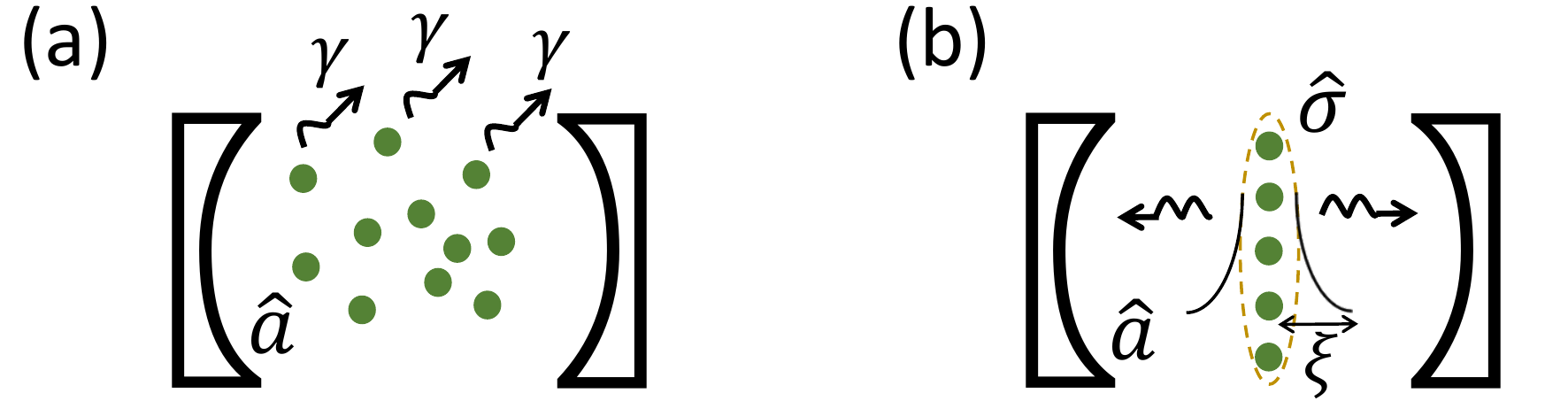}
    \caption{\small{
Many-atom cavity QED: interaction with the outside, non-confined modes.
(a) For a dilute cloud, atoms independently emit radiation to the non-confined modes at an individual-atom rate $\gamma$, typically approximated by that in free space.
(b) For an ordered atom array excited by the transversely confined cavity mode, the cooperative emission is directed into the cavity. Decay to outside, non-confined modes is therefore inhibited. The dispersive interactions between the atoms, dominated by near fields with longitudinal extent $\xi\ll$ cavity length, are essentially unaffected by the cavity structure.
    }} \label{fig2}
  \end{center}
\end{figure}

\section{System and model}
The system we consider is depicted in Fig. 3 and is comprised of the following parts.

\subsection{Atom array: internal and external degrees of freedom}
The atoms are modeled as two-level atoms with lowering operators $\hat{\sigma}_n$ ($n=1,...,N$) and corresponding dipole matrix element $d\mathbf{e}_d$ (with unit vector $\mathbf{e}_d$) trapped in a 2D lattice potential and spanning the $xy$ plane at $z=z_0$. For concreteness we assume that the $xy$ positions $\mathbf{r}_n^{\bot}$ form a square lattice with lattice spacing $a\lesssim\lambda$ (Fig. 1a, $\lambda$ being the wavelength of the cavity mode), but the results are straightforwardly generalized to other subwavelength lattices \cite{coop}. The array is taken to be effectively infinite, which is valid for small enough cross section of optical modes \cite{coop,AAMO}. Assuming very tight trapping in the $xy$ plane, we consider the motion of the atoms only along the longitudinal direction $z$, with a small-amplitude motion $\hat{z}_n$ around the identical equilibrium position $z_0$ (total longitudinal coordinate $z_0+\hat{z}_n$ for atom $n$). We assume harmonic longitudinal trapping potential with frequency $\omega_{\mathrm{m}}$ and zero-point motion $x_0=\sqrt{\hbar/(2M\omega_{\mathrm{m}})}$, $M$ being the atomic mass. The usual transformation to mechanical bosonic modes $\hat{b}_n$ is given by
\begin{equation}
\hat{b}_n=\frac{1}{2x_0}\left(\hat{z}_n+\frac{i}{M \omega_{\mathrm{m}}}\hat{p}_n\right),
\label{bn}
\end{equation}
with $\hat{p}_n$ the momentum of an atom $n$ along $z$.

\begin{figure}[t]
  \begin{center}
    \includegraphics[scale=0.45]{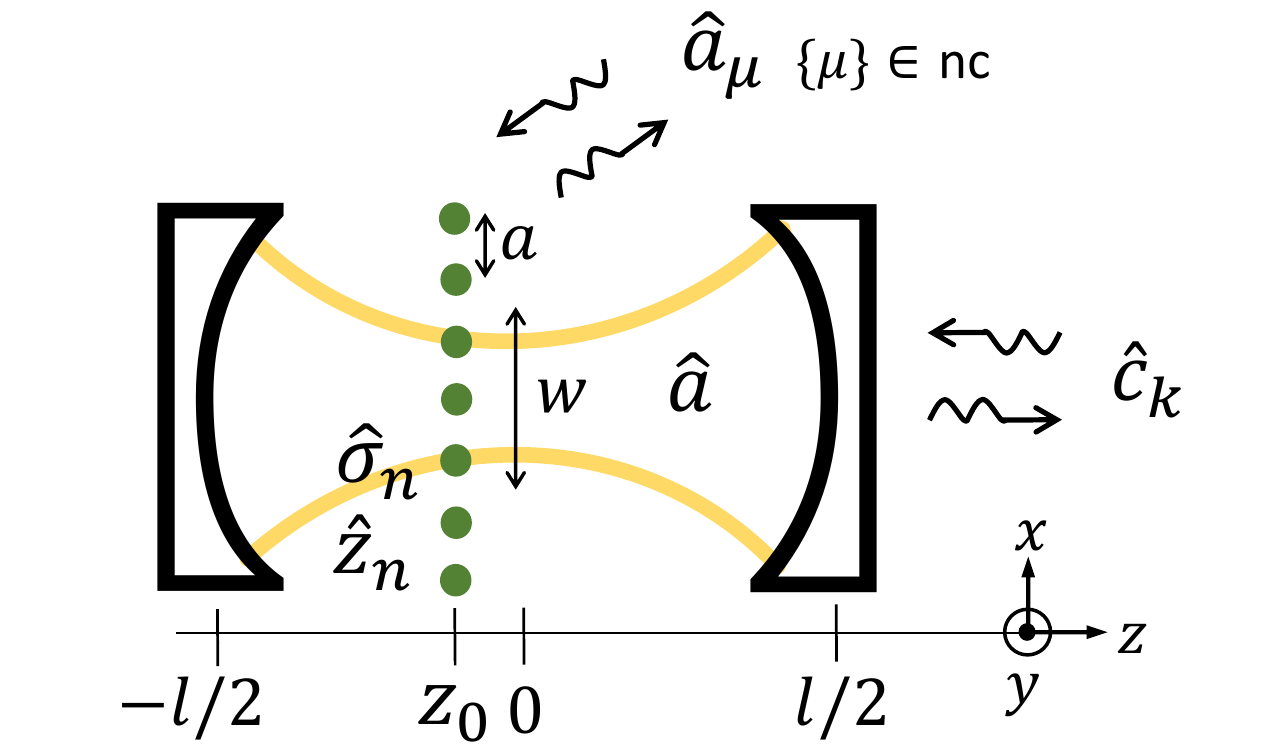}
    \caption{\small{
Quantum optical model of an atom array inside a cavity, see text in Sec. II.
    }} \label{fig3}
  \end{center}
\end{figure}

\subsection{Cavity mode}
Assuming that the typical cavity frequency spacing $\pi c/l$ is much larger than the typical frequencies of the relevant dynamics, we can consider only a single relevant cavity mode with a lowering boson operator $\hat{a}$ and a normalized mode profile, taken e.g. to be a Gaussian mode with frequency $\omega_c=cq=c2\pi/\lambda$ and waist $w$,
\begin{equation}
\mathbf{u}(\mathbf{r})=\sqrt{\frac{1}{\pi w^2l}} \frac{w}{\tilde{w}(z)}e^{-\frac{r_{\bot}^2}{\tilde{w}^2(z)}}\left(e^{i q \frac{r_{\bot}^2}{2 R(z)}}e^{i[qz-\psi(z)]}-\mathrm{c.c}\right)\mathbf{e}_d.
\label{u}
\end{equation}
Here $\mathbf{r}=(\mathbf{r}_{\bot},z)$ ($r_{\bot}=|\mathbf{r}_{\bot}|$), the polarization is $\mathbf{e}_d \bot \mathbf{e}_z$ and the beam parameters are given by $\tilde{w}(z)=w\sqrt{1+z^2/z_R^2}$, $R(z)=z(1+z_R^2/z^2)$ and $\psi(z)=\arctan(z/z_R)$, with $z_R=\pi w^2/\lambda$.
We further assume that the atomic longitudinal positions are well within the Rayleigh distance, $z_0, |\hat{z}_n|\ll z_R$, such that we may take $\psi(z)\rightarrow 0$, $R(z)\rightarrow\infty$ and $\tilde{w}(z)/w\rightarrow 1$ in  $\mathbf{u}(\mathbf{r}_n^{\bot},z_0+\hat{z}_n)$.

\subsection{Outside and non-confined modes}
The photon modes that are not confined by the cavity consist of two types. The first is the subset of modes  whose transverse profiles do not spatially match the cavity mirrors (e.g. being wider), including non-paraxial components. These transversely non-confined modes directly interact with the atoms and are referred to in the following as the \emph{non-confined modes}, with mode indices $\{\mu\}\in \mathrm{nc}$ (``nc" standing for non-confined) and corresponding annihilation operators, frequencies and mode functions $\hat{a}_{\mu}$, $\omega_{\mu}$ and $u_{\mu}(\mathbf{r})=\mathbf{u}_{\mu}(\mathbf{r})\cdot \mathbf{e}_d$, respectively (the latter already projected onto the atomic dipole orientation $\mathbf{e}_d$).
In addition, the finite mirror reflectivity allows for the out-coupling of the cavity mode $\hat{a}$ to a second type of outside modes, i.e. a 1D continuum of running paraxial modes with the same, transversely confined profile of the cavity modes. This 1D continuum, with mode functions $e^{ikz}/\sqrt{L}$ ($L \rightarrow \infty$), frequencies $\omega_k=|k|c$ and annihilation operators $\hat{c}_k$, merely gives rise to cavity damping and do not interact with the atoms directly.

\subsection{Hamiltonian}
Following the above considerations, the Hamiltonian of the full system reads,
\begin{eqnarray}
&&H=H_A+H_C+H_R+H_{AC}+H_{AR}+H_{CR},
\nonumber\\
&&H_A=\hbar\omega_a\sum_n \hat{\sigma}_n^{\dag}\hat{\sigma}_n +\sum_n\left[\frac{1}{2}M\omega_{\mathrm{m}}^2\hat{z}_n^2+\frac{\hat{p}_n^2}{2M}\right],
\nonumber\\
&&H_C=\hbar\omega_c\hat{a}^{\dag}\hat{a}+\hbar\left(\Omega e^{-i\omega_L t}\hat{a}^{\dag}+\mathrm{h.c.}\right),
\nonumber\\
&&H_R=\sum_{\mu\in\mathrm{nc}}\hbar\omega_{\mu}\hat{a}_{\mu}^{\dag}\hat{a}_{\mu}+\sum_k\hbar\omega_k\hat{c}_k^{\dag}\hat{c}_k,
\nonumber\\
&&H_{AC}=-\hbar \sum_n\left[i g_n\left(e^{iq(z_0+\hat{z}_n)}-\mathrm{h.c.}\right)\hat{a}\hat{\sigma}_n^{\dag}+\mathrm{h.c.}\right],
\nonumber\\
&&H_{AR}=-\hbar\sum_n\sum_{\mu\in\mathrm{nc}}\left[ig_{\mu}(\hat{\mathbf{r}}_n)\hat{a}_{\mu}(\hat{\sigma}_n^{\dag}+\hat{\sigma}_n)+\mathrm{h.c.}\right],
\nonumber\\
&&H_{CR}=\hbar\sqrt{\kappa_c\frac{c}{L}}\sum_k\left(\hat{c}_k\hat{a}^{\dag}+\mathrm{h.c.}\right),
\label{H}
\end{eqnarray}
with the dipole couplings,
\begin{eqnarray}
&&g_n=\sqrt{\frac{\omega_c}{2\varepsilon_0\hbar\pi w^2 l}}de^{-(r_n^{\bot}/w)^2},
\nonumber\\
&&g_{\mu}(\hat{\mathbf{r}}_n)=\sqrt{\frac{\omega_{\mu}}{2\varepsilon_0\hbar}}d u_{\mu}(\hat{\mathbf{r}}_n).
\label{gk}
\end{eqnarray}
Here, $H_A$ and $H_C$ are the atom and cavity Hamiltonians including the laser drive of the cavity mode (via the mirrors) with amplitude $\Omega$ and frequency $\omega_L$, and where $H_{AC}$ is their interaction. $H_R$ is the reservoir Hamiltonian consisting of the outside modes, with $H_{AR}$ the coupling of the atoms to the non-confined modes $\{\mu\}\in \mathrm{nc}$ and $H_{CR}$ the out-coupling of the cavity mode with couplings $\sqrt{\kappa_c c/L}$. The atomic coordinate $\hat{\mathbf{r}}_n=(\mathbf{r}^{\bot}_n,z_0+\hat{z}_n)$ appears as an operator due to the dynamical variable $\hat{z}_n$.

\section{Heisenberg-Langevin equations for non-saturated atoms: without motion}
At the first stage, we consider the atom array in the absence of motion, $\hat{z}_n,\hat{p}_n\rightarrow0$.
Following standard methods we begin by writing the Heisenberg equations of motion for the operators $\hat{a}$, $\hat{\sigma}_n$, $\hat{a}_{\mu}$ and $\hat{c}_k$. Solving formally for the reservoir modes $\hat{a}_{\mu}$ and $\hat{c}_k$ and inserting the solution into the equations for $\hat{\sigma}_n$ and $\hat{a}$, we assume weak couplings $g_{\mu}$ and the separation of timescales (Markov approximation) \cite{LEH,AAMO}
\begin{equation}
\frac{1}{\omega_L},\frac{L_a}{c}\ll \tau_s,
\label{mar}
\end{equation}
where $L_a=a\sqrt{N}$ is the size of the atomic system and $\tau_s$ the typical timescale of the envelope of the internal atomic states. The resulting Heisenberg-Langevin equations of motion are
\begin{eqnarray}
\dot{\tilde{a}}&=&\left(i\delta_c-\frac{\kappa_c}{2}\right)\tilde{a}-i\Omega-i2\sin(qz_0)\sum_n g^{\ast}_n\tilde{\sigma}_n+\hat{F}_c(t),
\nonumber\\
\dot{\tilde{\sigma}}_n&=&i\delta \tilde{\sigma}_n-i2g_n\sin(q z_0)\tilde{a}-\sum_{m}D_{nm}\tilde{\sigma}_m+\hat{F}_n(t),
\label{HL}
\end{eqnarray}
with the relevant operator envelopes and frequency detunings in a laser-rotating frame
\begin{equation}
\tilde{\sigma}_n=e^{i\omega_L t}\hat{\sigma}_n, \quad \tilde{a}=e^{i\omega_L t}\hat{a}, \quad  \delta=\omega_c-\omega_a, \quad \delta_c=\omega_L-\omega_c.
\label{det}
\end{equation}
In the equation for $\tilde{\sigma}_n$ the atomic dipolar excitations are linearized by taking $\hat{\sigma}^z_n=\hat{\sigma}^{\dag}_n\hat{\sigma}_n-\hat{\sigma}_n\hat{\sigma}^{\dag}_n\rightarrow -1$ (atoms far from saturation). The dipole-dipole interaction kernel,
\begin{equation}
D_{nm}=D(\mathbf{r}^{(0)}_n,\mathbf{r}^{(0)}_m),
\label{Dnm}
\end{equation}
with $\mathbf{r}^{(0)}_n=(\mathbf{r}_{\bot},z_0)$ denoting the equilibrium ordered-array positions, is mediated by the vacuum of the non-confined modes and is defined via
\begin{eqnarray}
D(\mathbf{r},\mathbf{r}')&=&\sum_{\mu\in\mathrm{nc}}\left[g_{\mu}(\mathbf{r})g^{\ast}_{\mu}(\mathbf{r}')\int_0^t dt'e^{-i(\omega_{\mu}-\omega_L)(t-t')}\right.
\nonumber\\
&&\left.-g^{\ast}_{\mu}(\mathbf{r})g_{\mu}(\mathbf{r}')\int_0^t dt'e^{i(\omega_{\mu}+\omega_L)(t-t')}\right],
\label{D1}
\end{eqnarray}
with the corresponding Langevin noise given by
\begin{equation}
\hat{F}_n(t)=-\sum_{\mu\in\mathrm{nc}} g_{\mu}(\mathbf{r}^{(0)}_n)\hat{a}_{\mu}(0)e^{-i(\omega_{\mu}-\omega_L)t}.
\label{Fn}
\end{equation}
Within the Markov approximation (\ref{mar}), one can typically show that the dipole-dipole kernel mediated by a complete set of electromagnetic modes is proportional to their corresponding  dyadic Green's function \cite{AAMO}. Here, the set of modes $\{\mu\}\in\mathrm{nc}$ does not span the full space. However, using a similar treatment, we find that $D(\mathbf{r},\mathbf{r}')$ from Eq. (\ref{D1}) can be written as
\begin{eqnarray}
D(\mathbf{r},\mathbf{r}')&=&-i\frac{3}{2}\gamma\lambda G_{\mathrm{nc}}(\mathbf{r},\mathbf{r}'),
\nonumber\\
G_{\mathrm{nc}}(\mathbf{r},\mathbf{r}')&=&\sum_{\mu\in\mathrm{nc}}\frac{u_\mu(\mathbf{r})u^{\ast}_\mu(\mathbf{r}')}{(\omega_\mu/c)^2-q^2},
\label{D2}
\end{eqnarray}
with $\gamma=q^3d^2/(3\pi\varepsilon_0\hbar)$ being the usual free-space spontaneous emission rate, and where $G_{\mathrm{nc}}(\mathbf{r},\mathbf{r}')$, referred to below as the \emph{non-confined Green's function}, is the component of the dyadic Green's function contributed by the non-confined modes $\mu\in$ nc.
Finally, for the reservoir modes due to the mirror out-coupling we have as usual
\begin{eqnarray}
&&\hat{F}_c(t)=-i\sqrt{\kappa_c c/L}\sum_k\hat{c}_k(0) e^{-i(\omega_k-\omega_L)t},
\nonumber\\
&&\langle\hat{F}_c(t)\hat{F}^{\dag}_c(t')\rangle=\kappa_c\delta(t-t').
\label{Fc}
\end{eqnarray}

\subsection{Collective dipole modes in free space}
As a reference case, consider now the situation of a fixed atomic array in the absence of the cavity (in free space). The only relevant equation of motion in Eq. (\ref{HL}) is then that for $\tilde{\sigma}$, with $\tilde{a}\rightarrow 0$,
\begin{equation}
\dot{\tilde{\sigma}}_n=-\frac{\gamma}{2}\tilde{\sigma}_n-\sum_{m\neq n}\tilde{\sigma}_m D^{\mathrm{fs}}_{nm}+\hat{F}^{\mathrm{fs}}_n(t),
\label{sig1}
\end{equation}
and where $\tilde{\sigma}_n=\hat{\sigma}_n e^{i\omega_a t}$ is now in an atom-rotating frame. The superscript ``fs" denotes that the dipole-dipole kernel (and corresponding Langevin force) is that mediated by free space modes (plane waves $\propto e^{i\mathbf{k}\cdot \mathbf{r}}$), for which $2\mathrm{Re}[D^{\mathrm{fs}}_{nn}]=\gamma$ is the usual spontaneous emission rate in free space and $\mathrm{Im}[D^{\mathrm{fs}}_{nn}]$ is a single-atom frequency shift which is neglected here. Moreover, $D^{\mathrm{fs}}(\mathbf{r}-\mathbf{r}')$ depends only on the separation $\mathbf{r}-\mathbf{r}'$ (see e.g. Appendix A). This motivates to move to 2D Fourier modes,
\begin{equation}
\hat{\sigma}_{\mathbf{k}_{\bot}}=\frac{1}{\sqrt{N}}\sum_{n=1}^{N}\hat{\sigma}_n e^{i\mathbf{k}_{\bot}\cdot\mathbf{r}^{\bot}_n},
\label{sigk}
\end{equation}
where $\mathbf{k}_{\bot}=(k_x,k_y)$ is inside the Brillouin zone $k_{x,y}\in\{-\pi/a,\pi/a\}$, for which Eq. (\ref{sig1}) becomes
\begin{equation}
\dot{\tilde{\sigma}}_{\mathbf{k}_{\bot}}=-\left[\frac{\gamma+\Gamma_{\mathbf{k}_{\bot}}}{2}+i\Delta_{\mathbf{k}_{\bot}}\right]\tilde{\sigma}_{\mathbf{k}_{\bot}}+\hat{F}^{\mathrm{fs}}_{\mathbf{k}_{\bot}}(t).
\label{sig1k}
\end{equation}
This result reveals that the normal modes of the atom array in free space and in the linear (non-saturated) regime are collective dipole modes $\hat{\sigma}_{\mathbf{k}_{\bot}}$ whose resonances are shifted by $\Delta_{\mathbf{k}_{\bot}}$ (cooperative shift) and broadened by $\Gamma_{\mathbf{k}_{\bot}}$ (cooperative emission) with respect to the individual-atom resonance $\omega_a$ and $\gamma$ \cite{coop}. Furthermore, the cooperative emission of dipole mode $\mathbf{k}_{\bot}$ is directed solely to photon modes with the same $\mathbf{k}_{\bot}$ in the transverse direction (Fig. 1b, for $|\mathbf{k}_{\bot}|<2\pi/\lambda$) \cite{note2}. Due to the translational symmetry, the cooperative resonances are diagonal in $\mathbf{k}_{\bot}$ and given by
\begin{eqnarray}
&&\Gamma_{\mathbf{k}_{\bot}}=2\sum_{n\neq 1}e^{-i\mathbf{k}_{\bot}\cdot\mathbf{r}^{\bot}_n}\mathrm{Re}[D^{\mathrm{fs}}_{n1}],
\nonumber\\
&&\Delta_{\mathbf{k}_{\bot}}=\sum_{n\neq 1}e^{-i\mathbf{k}_{\bot}\cdot\mathbf{r}^{\bot}_n}\mathrm{Im}[D^{\mathrm{fs}}_{n1}],
\nonumber\\
&&\Delta\equiv\Delta_{\mathbf{k}_{\bot}=0}, \quad \Gamma\equiv\Gamma_{\mathbf{k}_{\bot}=0}=\frac{3}{4\pi}\frac{\lambda^2}{a^2}\gamma-\gamma,
\label{Dfs}
\end{eqnarray}
where here $n=1$ denotes the atom located at the origin of an effectively infinite array. These cooperative resonances are discussed in depth in Ref. \cite{coop}, wherein an analytic expression for $\Gamma_{\mathbf{k}_{\bot}}$ is found for any $\mathbf{k}_{\bot}$.

\section{Generic treatment of the non-confined Green's function}
The effect of the non-confined modes on the dynamics is encoded in the non-confined Green's function $G_{\mathrm{nc}}$, via the dipole-dipole kernel $D$ that appears in the dynamical equations. The next step, which is the topic of this section,  is therefore to find a physically motivated approximation for its evaluation. Before we turn to a more technical discussion, let us first briefly summarize the theoretical challenge in treating the non-confined Green's function $G_{\mathrm{nc}}$ and the approach taken here.

\emph{The problem.---} In general, in order to find $G_{\mathrm{nc}}$ and $D$ from Eq. (\ref{D2}), one has to solve for the non-confined spatial modes $u_{\mu}$ which depend on the cavity structure and geometry and are hence system specific and typically difficult to find \cite{noteWG}. The typical generic approximation in cavity and waveguide QED is then to approximate $D$ by that in free space $D^{\mathrm{fs}}$. E.g. for a single atom in a cavity/waveguide, this may give a good approximation to the decay rate to outside, non-confined modes. However, when collective effects mediated between the atoms by these non-confined modes become important, this approximation may lead to completely wrong results. E.g. in our case, a transversely confined collective dipole mode of the array scatters light only into the equivalent confined cavity modes; hence, its decay rate due to outside modes is expected to vanish, in contrast to the prediction of the typical approximation $D\approx D^{\mathrm{fs}}$ which yields the non-vanishing decay from Eq. (\ref{Dfs}).

\emph{Our approach.---} We wish to refine the typical, free-space generic approximation for $D$ by considering the following physically motivated modifications. For a collective dipole mode whose spatial profile matches that of a cavity mode, we expect: (a) Its decay rate to non-confined modes vanishes, $\mathrm{Re}[D]=0$; (b) The frequency shift due to non-confined modes, $\mathrm{Im}[D]$, does not vanish, and is treated using the usual approximation $D^{\mathrm{fs}}$. Since the dispersive shift is dominated mostly be near fields, the latter is expected to be a good approximation for atoms positioned not too close to the cavity mirrors, as further explained below.

\subsection{Physical considerations and approximation}
In order to get more insight into our approach, imagine that we span the space of in-plane atomic dipole positions $\mathbf{r}^{\bot}_n$ ($n=1,...,N$) with a set of $N$ orthonormal modes $u_{\alpha,n}$ ($\alpha=1,...,N$), the first of which  ($\alpha=1$) has the transverse profile of the cavity mode, $u_{1,n}=\sqrt{a^2/\pi w^2}e^{-|\mathbf{r}^{\bot}_n|^2/w^2}$. Projecting  Eq. (\ref{HL}) for $\tilde{\sigma}_n$ onto the cavity-profile mode $\alpha=1$, we have
\begin{eqnarray}
\dot{\tilde{\sigma}}=i\delta \tilde{\sigma}-ig_c \sin(qz_0)\tilde{a}-\sum_{\alpha}\tilde{\sigma}_{\alpha}D_{1,\alpha}+\hat{F},
\label{28}
\end{eqnarray}
with $\tilde{\sigma}=\tilde{\sigma}_{\alpha=1}$ and
\begin{eqnarray}
&&\tilde{\sigma}_{\alpha}=\sum_n u^{\ast}_{\alpha,n}\tilde{\sigma}_n, \quad g_c=\sqrt{\frac{\omega_c}{2\varepsilon_0\hbar l a^2}}d,
\nonumber\\
&&D_{1,\alpha}=\sum_n\sum_m u_{1,n}D_{nm}u^{\ast}_{\alpha,m}.
\label{28a}
\end{eqnarray}
Here $\hat{F}=\sum_n u^{\ast}_{\alpha,n}\hat{F}_n$ and for the evaluation of $g_c$ we used $w\gg a$ to convert sums to integrals, $a^2\sum_n\rightarrow \int d \mathbf{r}_{\bot}$.

As usual, $D_{1,\alpha}$ can be understood as an effective, non-Hermitian dipole-dipole Hamiltonian that couples the collective atomic-dipole modes $\tilde{\sigma}_{\alpha}$. Therefore, $\mathrm{Im} [D_{1,\alpha}]$ describes the coherent, Hamiltonian process mediated by virtual photons from the non-confined modes $\mu\in\mathrm{nc}$, whereas $\mathrm{Re} [D_{1,\alpha}]$ describes the collective atomic damping (emission) to these non-confined modes.
To understand the properties of $D_{1,\alpha}$, it is sufficient to consider the field emanating from an array of classical oscillating dipoles wherein the polarization amplitude of a dipole at array position $\mathbf{r}^{\bot}_{n}$ is proportional to the corresponding confined-mode amplitude $u_{1,n}=u(\mathbf{r}^{\bot}_n)$ , and in the absence of the cavity mirrors (see the supplement of Ref. \cite{coop} for details).
Within this picture, the array effectively acts as a grating for an exciting field with the cavity transverse profile $u(\mathbf{r}_{\bot})$, denoted $u(\mathbf{k}_{\bot})$ in 2D spatial Fourier representation. Therefore, in the emanating field, each $\mathbf{k}_{\bot}$ that is supported by $u(\mathbf{k}_{\bot})$ will be shifted by the reciprocal lattice vectors of the array, $\mathbf{k}_{\bot}\rightarrow\mathbf{k}_{\bot}+\mathbf{Q}_{m_x,m_y}$ with $\mathbf{Q}_{m_x,m_y}=2\pi/a(m_x\mathbf{e}_x+m_y\mathbf{e}_y)$ ($m_{x,y}=0,\pm 1, \pm 2,...$), and with corresponding longitudinal wavenumbers
\begin{eqnarray}
k^{m_x,m_y}_z=\sqrt{\left(\frac{2\pi}{\lambda}\right)^2-\left|\mathbf{k}_{\bot}+\frac{2\pi}{a}(m_x\mathbf{e}_x+m_y\mathbf{e}_y)\right|^2}.
\label{29e}
\end{eqnarray}
Since the emanating field can have a propagating, radiated part, for which $k_z$ is real, and a non-propagating, near-field part, with $k_z$ imaginary, we may deduce the properties of $D_{1,\alpha}$ as follows.
\\
\emph{(a) Emission process} $\mathrm{Re}[D_{1,\alpha}]$: Considering the narrow-band spatial frequency content of the paraxial cavity mode $u(\mathbf{k}_{\bot})$ around $\mathbf{k}_{\bot}=0$, and assuming $a<\lambda$, we observe that $k_z$ is real only for $m_x=m_y=0$  (zeroth diffraction order) \cite{note1}, so that the radiated, propagating field has the original cavity mode profile $u(\mathbf{k}_{\bot})$, and will be contained in the cavity (Fig. 2b). Hence, no emission and damping to the non-confined modes is possible, $\mathrm{Re}[D_{1,\alpha}]=0$.
\\
\emph{(b) Coherent dipole-dipole shifts} $\mathrm{Im}[D_{1,\alpha}]$: The non-radiative field components, with imaginary $k_z$, give rise to the collective dipole-dipole shifts such as $\Delta_{\mathbf{k}_{\bot}}$ in free space (see supplement of Ref. \cite{coop}). For $a<\lambda$, $k_z$ is imaginary for $m_x,m_y\neq0,0$ as explained above, leading to near fields with a longitudinal extent $\xi\sim 1/|k_z|\lesssim a$ away from the array. Then, for an array at a distance larger than $\lambda>\xi\sim a$ from the cavity mirrors, these near fields and the coherent dipole-dipole interactions associated with them are not affected by the cavity structure (Fig. 2b). Therefore, the dispersive dipole-dipole effects should be well approximated by those in free space, leading to $\mathrm{Im}[D_{nm}]\approx \mathrm{Im}[D^{\mathrm{fs}}_{nm}]$ and more specifically to $\mathrm{Im}[D_{1,\alpha}]\approx \mathrm{Im}[D^{\mathrm{fs}}_{1,\alpha}]$.

To conclude, for an array with $a<\lambda$ and a cavity length $l\gg\lambda$, the above physical considerations lead to the generic approximation
\begin{subequations}
\begin{align}
&\mathrm{Re}[D_{1,\alpha}]=0,
\label{30a}
\\
&\mathrm{Im}[D_{1,\alpha}]\approx \mathrm{Im}[D^{\mathrm{fs}}_{1,\alpha}], \quad \mathrm{Im}[D_{nm}]\approx \mathrm{Im}[D^{\mathrm{fs}}_{nm}].
\label{30b}
\end{align}
\label{30}
\end{subequations}
We note that the same considerations should hold for any cavity-confined mode.

\subsection{Generalization of the generic approximation}
For the cavity QED description of a motionless atom array, it is enough to consider the conditions (\ref{30}) which are stated in terms of the Green's function (dipole-dipole kernel) at fixed array positions, $D_{nm}=D(\mathbf{r}^{(0)}_n,\mathbf{r}^{(0)}_m)$, as illustrated in the next subsection. However, when treating the small motion of atoms around the equilibrium array positions, the spatial derivatives of $D(\mathbf{r},\mathbf{r}')$ around these positions are also important (see Sec. V below), and a more general statement on this Green's function is required. To this end, and since the approximation (\ref{30}) should hold for any mode whose transverse profile matches that of the confined-cavity modes, we can write this approximation in a more general from as
\begin{subequations}
\begin{align}
&D(\mathbf{r},\mathbf{r}')\approx D^{\mathrm{fs}}(\mathbf{r},\mathbf{r}')-D^{\mathrm{c}}(\mathbf{r},\mathbf{r}'),
\label{31a}
\\
&\mathrm{Im}[D(\mathbf{r},\mathbf{r}')]\approx\mathrm{Im}[D^{\mathrm{fs}}(\mathbf{r},\mathbf{r}')].
\label{31b}
\end{align}
\label{31}
\end{subequations}
The approximation (\ref{31a}) is a refinement of the typical approximation for the non-confined Green's function $D\approx D^{\mathrm{fs}}$ wherein we project out the contribution due to the subspace of all photon modes whose transverse mode profiles match the cavity mirrors and are transversely-confined by them. This is performed by subtracting
$D^c\propto \sum_{\mu\in\mathrm{c}}\frac{u_\mu(\mathbf{r})u^{\ast}_\mu(\mathbf{r}')}{(\omega_\mu/c)^2-q^2}$ from $D^{\mathrm{fs}}$, where $\{\mu\}\in\mathrm{c}$ is a set of modes that span this transversely confined subspace, e.g. the paraxial Hermite-Gauss modes (Appendix A). This construction guarantees that condition (\ref{30}a) is satisfied, whereas the further approximation (\ref{31b}) corresponds to that of (\ref{30}b).

\subsection{Atom-cavity dynamics in the absence of motion}
As a first illustration, we consider the resulting atom-cavity coupled dynamics in the absence of motion. Starting with Eq. (\ref{28}), denoting $\Delta_{1\alpha}=\mathrm{Im}[D_{1,\alpha}]$, we have
\begin{eqnarray}
\dot{\tilde{\sigma}}=i\delta \tilde{\sigma}-ig_c \sin(qz_0)\tilde{a}-i\sum_{\alpha\neq 1}\tilde{\sigma}_{\alpha}\Delta_{1\alpha}\tilde{\sigma}_{\alpha}-i\Delta_{11}\tilde{\sigma},
\label{F27}
\end{eqnarray}
where we used $\mathrm{Re}[D_{1,\alpha}]=0$, noting that the Langevin noise vanishes accordingly, as per the fluctuation-dissipation theorem. From Eq. (\ref{30}b) we have that the matrix $\Delta_{\alpha \alpha'}$ is identical to that in free space and is therefore diagonal in the lattice momentum basis with dispersion relation (eigenvalues) $\Delta_{\mathbf{k}_{\bot}}$ from Eq. (\ref{Dfs}). For a dispersion relation sufficiently flat around $\mathbf{k}_{\bot}=0$, such that
$\Delta_{\mathbf{k}_{\bot}}\approx\Delta_{\mathbf{k}_{\bot}=0}\equiv \Delta$ for paraxial wavevectors $\mathbf{k}_{\bot}$, we have that the matrix $\Delta_{\alpha \alpha'}$ projected to the paraxial subspace, is proportional to the identity matrix. Since the cavity mode is paraxial, we take $\Delta_{11}\approx \Delta$ and $\Delta_{1\alpha}\approx 0$ \cite{note}, so that Eq. (\ref{F27}) together with the Eq. (\ref{HL}) for $\tilde{a}$ yield
\begin{eqnarray}
&&\dot{\tilde{a}}=\left(i\delta_c-\frac{\kappa_c}{2}\right)\tilde{a}-ig_c\tilde{\sigma}\sin(qz_0)-i\Omega+\hat{F}_c,
\nonumber\\
&&\dot{\tilde{\sigma}}=i(\delta-\Delta)\tilde{\sigma}-ig_c\sin(qz_0)\tilde{a}.
\label{F31}
\end{eqnarray}
These equations describe the dynamics of two coupled oscillators, a damped cavity mode $\tilde{a}$, and a collective dipole mode $\tilde{\sigma}$ (non-saturated). The latter inherits the spatial profile of the cavity mode [Eq. (\ref{28a})], so that its cooperative emission to the non-confined modes vanishes and it appears as an undamped oscillator.

\section{Heisenberg-Langevin equations including motion}
The generic approximation, Eq. (\ref{30}), was obtained by considering a fixed (motionless) ordered array. However, Eq. (\ref{30}) yields a general statement on the non-confined Green's function, which can be used also when small-amplitude motion around the equilibrium array positions is considered. In this section, we first restore the motional degrees of freedom into the atom-cavity equations of motion, which are then used to analyze optomechanical effects in later sections.

Considering also the motional degrees of freedom with operators $\hat{z}_n$ and $\hat{p}_n$ and using similar methods to those described in Sec. III, the Heisenberg-Langevin equations from Eq. (\ref{HL}) are generalized to include also the equation for $\hat{p}_n$, with an additional condition for the Markov-approximation,  $\frac{1}{\omega_L},\frac{L_a}{c}\ll \lambda/\dot{z}_n$, yielding
\begin{eqnarray}
\dot{\tilde{\sigma}}_n&=&i\delta \tilde{\sigma}_n-i2g_n\sin[q(z_0+\hat{z}_n)]\tilde{a}-\sum_{m}\tilde{\sigma}_m D(\hat{\mathbf{r}}_n,\hat{\mathbf{r}}_m)
\nonumber\\
&&+\hat{F}_n(t),
\nonumber\\
\dot{\hat{p}}_n&=&-M\omega_{\mathrm{m}}^2\hat{z}_n-2\hbar q\cos[q(z_0+\hat{z}_n)]\left[g_n\tilde{a}\tilde{\sigma}^{\dag}_n+\mathrm{h.c.}\right]
\nonumber\\
&&-\sum_{m}\left[A(\hat{\mathbf{r}}_n,\hat{\mathbf{r}}_m)\tilde{\sigma}_n^{\dag}\tilde{\sigma}_m+\mathrm{h.c.}\right] +\hat{f}_n(t),
\nonumber\\
\dot{\tilde{a}}&=&\left(i\delta_c-\frac{\kappa_c}{2}\right)\tilde{a}-i\Omega-\sum_n g^{\ast}_n\left(e^{iq(z_0+\hat{z}_n)}-\mathrm{h.c.}\right)\tilde{\sigma}_n
\nonumber\\
&&+\hat{F}_c(t),
\nonumber\\
\label{24a}
\end{eqnarray}
and $\dot{\hat{z}}_n=\hat{p}_n/M$, with $\hat{\mathbf{r}}_n=(\mathbf{r}_n^{\bot},z_0+\hat{z}_n)$, noting that the Langevin noise $\hat{F}_n(t)$ from Eq. (\ref{Fn}) now contains the operator $\hat{\mathbf{r}}_n$ instead of the equilibrium position $\mathbf{r}^{(0)}_n$.

In the equation for $\hat{p}_n$ we identify the forces between atoms resulting from the dipole-dipole interactions mediated by non-confined modes, along with the corresponding Langevin force
\begin{eqnarray}
&&A(\mathbf{r},\mathbf{r}')=-i\hbar \frac{\partial}{\partial z} D(\mathbf{r},\mathbf{r}'),
\nonumber\\
&&\hat{f}_n=\tilde{\sigma}_n^{\dag}\left[-i\hbar \frac{\partial}{\partial z_n} \hat{F}_n(t,z_n)\right]_{z_n=\hat{z}_n}+\mathrm{h.c.}.
\label{A}
\end{eqnarray}

\subsection{Small-amplitude motion}
We now wish to expand the dynamical equations around the equilibrium positions, assuming that the motion $\hat{z}_n$ around $z_0$ is much smaller than a cavity wavelength,
\begin{equation}
q \hat{z}_n\ll 1 \quad \Leftrightarrow \quad \hat{z}_n\ll \lambda/(2\pi).
\label{52}
\end{equation}
A consistent description of the dynamical equations (\ref{24a}) is obtained by expanding the equations for $\dot{\tilde{\sigma}}_n$ and $\dot{\tilde{a}}$ up to second order in $q \hat{z}_n$ while expanding that for $\dot{\hat{p}}_n$ up to first order (since the latter is an equation of motion for the conjugate variable of $\hat{z}_n$), yielding
\begin{eqnarray}
\dot{\tilde{\sigma}}_n&=&i\delta \tilde{\sigma}_n-i2g_n\left[\sin(q z_0)\left(1-\frac{q^2\hat{z}_n^2}{2}\right)+\cos(q z_0)q\hat{z}_n\right]\tilde{a}
\nonumber\\
&&-\sum_{m}\tilde{\sigma}_m \left(D_{nm}+\hat{J}_{nm}\right)+\hat{F}_n(t),
\nonumber\\
\dot{\hat{p}}_n&=&-M\omega_{\mathrm{m}}^2\hat{z}_n
-\sum_{m}\left[\hat{A}_{nm}\tilde{\sigma}_n^{\dag}\tilde{\sigma}_m+\mathrm{h.c.}\right] +\hat{f}_n(t)
\nonumber\\
&&-2\hbar q\left[\cos(qz_0)-\sin(qz_0)q\hat{z}_n)\right]\left[g_n\tilde{a}\tilde{\sigma}^{\dag}_n+\mathrm{h.c.}\right]
\nonumber\\
\dot{\tilde{a}}&=&\left(i\delta_c-\frac{\kappa_c}{2}\right)\tilde{a}-i\Omega+\hat{F}_c(t)
\nonumber\\
&&-i2\sum_n g^{\ast}_n\left[\sin(qz_0)\left(1-\frac{q^2\hat{z}_n^2}{2}\right)+\cos(qz_0)q\hat{z}_n\right]\tilde{\sigma}_n.
\nonumber\\
\label{27}
\end{eqnarray}
Here we used the formal expansion of the dipole-dipole kernels around the equilibrium positions $\mathbf{r}_n^{(0)}=(\mathbf{r}_n^{\bot},z_0)$,
\begin{eqnarray}
D(\hat{\mathbf{r}}_n,\hat{\mathbf{r}}_m)&\approx&D_{nm}+\hat{J}_{nm},
\nonumber\\
\hat{J}_{nm}&=&\sum_{s=n,m}\frac{\partial}{\partial z_s}\left.D(\mathbf{r}_n,\mathbf{r}_m)\right|_{\mathbf{r}_{n,m}^{(0)}}\hat{z}_s
\nonumber\\
&&+\frac{1}{2}\sum_{s,s'=n,m}\frac{\partial}{\partial z_s}\frac{\partial}{\partial z_s'}\left.D(\mathbf{r}_n,\mathbf{r}_m)\right|_{\mathbf{r}_{n,m}^{(0)}}\hat{z}_s\hat{z}_{s'},
\nonumber\\
\label{26D}
\end{eqnarray}
with $D_{nm}$ from (\ref{Dnm}), and
\begin{eqnarray}
\hat{A}_{nm}\approx A(\mathbf{r}_n^{(0)},\mathbf{r}_m^{(0)})+\sum_{s=n,m}\left.\frac{\partial}{\partial z_s}A(\mathbf{r}_n,\mathbf{r}_m)\right|_{\mathbf{r}_{n,m}^{(0)}}\hat{z}_s.
\label{26A}
\end{eqnarray}
The approximation $D\approx D^{\mathrm{fs}}-D^c$ from (\ref{31a}) allows us to slightly simplify the above expressions. Recalling our assumption of an array placed very close to the focus of the cavity modes ($z_0\ll z_R=\pi w^2/\lambda$), the confined Green's function $D^c(\mathbf{r},\mathbf{r}')$ is approximately symmetric around the equilibrium position $z_0$ (Appendix A). Since this is the case also for the free-space kernel $D^{\mathrm{fs}}(\mathbf{r},\mathbf{r}')$, the first-order in the expansion for $D$ (and zeroth order for $A$) vanishes, yielding (Appendix A)
\begin{eqnarray}
&&\hat{J}_{nm}\approx \frac{1}{2}D''_{nm}(\hat{z}_n-\hat{z}_m)^2, \:\: D''_{nm}=\left.\frac{\partial^2}{\partial z^2}D(\mathbf{r},\mathbf{r}^{(0)}_m)\right|_{\mathbf{r}=\mathbf{r}^{(0)}_n},
\nonumber\\
&& \hat{A}_{nm}\approx -i\hbar D''_{nm}(\hat{z}_n-\hat{z}_m).
\label{33c}
\end{eqnarray}
We note that the transition from Eqs. (\ref{26D}) and (\ref{26A}) to Eqs. (\ref{33c}) is the only step performed in this work in which the generalized approximation (\ref{31}) [instead of the more restricted approximation (\ref{30})] was strictly required.

Finally, the Langevin forces $\hat{F}_n$ and $\hat{f}_n$ in Eq (\ref{27}) should also be understood as the expansions around $z_0$ up to second and first order in $\hat{z}_n$, respectively, of the corresponding expressions in Eqs. (\ref{Fn}) and (\ref{A}).

\section{Large atom-cavity detuning: optomechanical coupling}
The current and subsequent sections are devoted to the formulation of the optomechanics of an atom-array inside a cavity. We consider the case where the atom array acts as a partially reflecting membrane whose motion modifies the cavity resonance, in analogy to typical cavity optomechanical setups \cite{AKM} (Fig. 4). This situation can be reached when the internal electronic degrees of freedom of the atoms are eliminated, resulting in an optomechanical system of atomic motion coupled to cavity light.

To eliminate the internal states, a separation of timescales should be identified, which is facilitated by transforming the equation for $\tilde{\sigma}_n$ in (\ref{27}) to $\mathbf{k}_{\bot}$-space using Eq. (\ref{sigk}) and by applying the approximation (\ref{30b}) for the non-confined Green's function, yielding,
\begin{equation}
\dot{\tilde{\sigma}}_{\mathbf{k}_{\bot}}=i(\delta-\Delta_{\mathbf{k}_{\bot}})\tilde{\sigma}_{\mathbf{k}_{\bot}}
-\sum_{\mathbf{k}'_{\bot}}\tilde{\sigma}_{\mathbf{k}'_{\bot}}\frac{\hat{\Gamma}_{\mathbf{k}_{\bot}\mathbf{k}'_{\bot}}}{2}
+\hat{B}_{\mathbf{k}_{\bot}}+\hat{F}_{\mathbf{k}_{\bot}}.
\label{35}
\end{equation}
Here,
\begin{eqnarray}
&&\hat{\Gamma}_{\mathbf{k}_{\bot}\mathbf{k}'_{\bot}}=\gamma_{\mathbf{k}_{\bot}\mathbf{k}'_{\bot}}+2\hat{J}_{\mathbf{k}_{\bot}\mathbf{k}'_{\bot}},
\nonumber\\
&&\gamma_{\mathbf{k}_{\bot}\mathbf{k}'_{\bot}}=\frac{1}{N}\sum_n\sum_m e^{-i\mathbf{k}_{\bot}\cdot\mathbf{r}^{\bot}_n}2\mathrm{Re}[D_{nm}]e^{i\mathbf{k}'_{\bot}\cdot\mathbf{r}^{\bot}_m},
\nonumber\\
&&\hat{J}_{\mathbf{k}_{\bot}\mathbf{k}'_{\bot}}=\frac{1}{N}\sum_n\sum_m e^{-i\mathbf{k}_{\bot}\cdot\mathbf{r}^{\bot}_n}\hat{J}_{nm}e^{i\mathbf{k}'_{\bot}\cdot\mathbf{r}^{\bot}_m},
\label{33a}
\end{eqnarray}
and
\begin{eqnarray}
\hat{B}_{\mathbf{k}_{\bot}}&=&-i2\frac{1}{\sqrt{N}}\sum_n e^{-i\mathbf{k}_{\bot}\cdot\mathbf{r}^{\bot}_n}g_n
\nonumber\\
&\times&\left[\sin(qz_0)\left(1-\frac{q^2\hat{z}_n^2}{2}\right)+q\hat{z}_n\cos(qz_0)\right]\tilde{a},
\nonumber\\
\hat{F}_{\mathbf{k}_{\bot}}&=&\frac{1}{\sqrt{N}}\sum_n e^{-i\mathbf{k}_{\bot}\cdot\mathbf{r}^{\bot}_n}\hat{F}_n.
\label{34}
\end{eqnarray}

\subsection{Large-detuning approximation: elimination of internal states}
We begin by assuming that the detuning between the cavity and atomic resonances (including the cooperative shift of the latter) is much greater than all relevant frequencies of the dynamical variables of interest,
\begin{eqnarray}
|\delta-\Delta_{\mathbf{k}_{\bot}}|\sim |\delta-\Delta|\gg \tau_d^{-1}\sim \dot{\hat{b}}_n/\hat{b}_n,\dot{\tilde{a}}/\tilde{a},\Gamma+\gamma,
\label{37}
\end{eqnarray}
with $\hat{b}_n$ the bosonic motion operator [Eq. (\ref{bn})], and recalling $\Delta=\Delta_{\mathbf{k}_{\bot}=0}$ and $\Gamma=\Gamma_{\mathbf{k}_{\bot}=0}$ from Eq. (\ref{Dfs}). To adiabatically eliminate the internal degrees of freedom we then move to a coarse-grained dynamical picture with time resolution $T$ satisfying $|\delta-\Delta_{\mathbf{k}_{\bot}}|^{-1}\ll T\ll\tau_d$. We formally solve Eq. (\ref{35}) around an ``initial" time $t$ up to some time $t_1$ within the coarse-graining time-resolution, $|t_1-t|<T$,
\begin{eqnarray}
\tilde{\sigma}_{\mathbf{k}_{\bot}}(t_1)&=&e^{i(\delta-\Delta_{\mathbf{k}_{\bot}})(t_1-t)}\tilde{\sigma}_{\mathbf{k}_{\bot}}(t)
\nonumber\\
&+&\int_{t}^{t_1} dt' e^{i(\delta-\Delta_{\mathbf{k}_{\bot}})(t_1-t')}\left[\hat{F}_{\mathbf{k}_{\bot}}(t')+\hat{B}_{\mathbf{k}_{\bot}}(t)\right]
\nonumber\\
&-&\sum_{\mathbf{k}'_{\bot}}\int_{t}^{t_1} dt' e^{i(\delta-\Delta_{\mathbf{k}_{\bot}})(t-t')}\tilde{\sigma}_{\mathbf{k}'_{\bot}}(t')\frac{1}{2}\hat{\Gamma}_{\mathbf{k}_{\bot}\mathbf{k}_{\bot}}(t),
\nonumber\\
\label{38}
\end{eqnarray}
noting that $\hat{B}_{\mathbf{k}_{\bot}}(t)$ and $\hat{\Gamma}_{\mathbf{k}_{\bot}\mathbf{k}_{\bot}}(t)$ can be evaluated at $t$ since they vary at a timescale $\sim\tau_d\gg T$ and are therefore approximately unchanged during the time resolution $T$. Moreover, we note that the third term is of order
$T \hat{\Gamma}_{\mathbf{k}_{\bot}\mathbf{k}_{\bot}}$ and is therefore small since
$\hat{\Gamma}_{\mathbf{k}_{\bot}\mathbf{k}_{\bot}}\lesssim \gamma+\Gamma \ll T^{-1}$ (Eq. \ref{37}).
We can then treat this term as a perturbation and to lowest order replace $\tilde{\sigma}_{\mathbf{k}_{\bot}}(t')$ that appears in it by the ``free" solution given by the first two terms of Eq. (\ref{38}). Moving to coarse-grained variables, $\tilde{\sigma}_{\mathbf{k}_{\bot}}(t)\rightarrow (1/T)\int_{t}^{t+T}dt_1\tilde{\sigma}_{\mathbf{k}_{\bot}}(t_1)$ and dropping terms of order $T^{-1}/(\delta-\Delta_{\mathbf{k}_{\bot}})$, we finally obtain the coarse-grained steady-state solution for the internal state in the form
\begin{eqnarray}
\tilde{\sigma}_{\mathbf{k}_{\bot}}(t)=\hat{W}_{\mathbf{k}_{\bot}}\tilde{a}+\widehat{\delta\sigma}_{\mathbf{k}_{\bot}}(t).
\label{53a}
\end{eqnarray}
The coefficient $\hat{W}_{\mathbf{k}_{\bot}}$ is an operator that contains the small parameter $q\hat{z}_n$ up to second order, whereas $\widehat{\delta\sigma}_{\mathbf{k}_{\bot}}$ is a noise term arising from the Langevin operator $\hat{F}_{\mathbf{k}_{\bot}}$ associated with the non-confined modes. The explicit expressions for both $\hat{W}_{\mathbf{k}_{\bot}}$ and $\widehat{\delta\sigma}_{\mathbf{k}_{\bot}}$ are given in Appendix B.

\subsection{Equation for cavity mode}
As a first step for an optomechanical description of the system, we eliminate the internal degrees of freedom from the equation for the cavity mode by inserting Eq. (\ref{53a}) into Eq. (\ref{27}) for $\tilde{a}$. The resulting equation for the cavity mode is simplified by using the paraxial nature of the cavity mode, with waist satisfying $w\gg\lambda >a$, along with the condition from Eq. (\ref{30a}), obtaining (Appendix B 1)
\begin{eqnarray}
\dot{\tilde{a}}&=&\left[i(\delta_c-\Delta_{\mathrm{AC}})-\frac{\kappa_c+\hat{K}_{sc}}{2}\right]\tilde{a}-ig\left(\hat{b}+\hat{b}^{\dag}\right)\tilde{a}
\nonumber\\
&&-i\Omega+\hat{F}_c(t)+\hat{F}_{sc}(t),
\label{72}
\end{eqnarray}
with the optomechanical coupling $g$ and the dispersive, atom-induced cavity shift $\Delta_{\mathrm{AC}}$ given by
\begin{eqnarray}
&&g=\sin(2qz_0)\eta \bar{g}, \quad \bar{g}=\frac{c}{l}\frac{\gamma}{\delta-\Delta}\sqrt{N_a}\frac{3}{q^2 w^2},
\nonumber\\
&&\Delta_{\mathrm{AC}}=\sin^2(qz_0)\frac{c}{l}\frac{\gamma+\Gamma}{\delta-\Delta},
\label{71}
\end{eqnarray}
where
\begin{eqnarray}
\eta=q x_0\ll 1, \quad N_a=\pi \frac{w^2}{a^2},
\label{eta}
\end{eqnarray}
are, respectively, the Lamb-Dicke parameter (with $x_0$ the zero-point motion from Sec. II A) and  the number of atoms within the cavity-mode waist. The relevant collective mechanical mode coupled to the cavity mode inherits the intensity profile of the latter,
\begin{eqnarray}
\hat{b}=\sum_{n} V_n^0 \hat{b}_n, \quad V_n^0=\frac{2}{\sqrt{\pi}}\frac{a}{w}e^{-2(r_n^{\bot}/w)^2},
\label{73}
\end{eqnarray}
with $\sum_n (V_n^0)^2= 1$ (using $w\gg a$). The effective damping operator $\hat{K}_{sc}$ and Langevin noise $\hat{F}_{sc}(t)$ are discussed below.

\subsection{Towards standard cavity optomechanics}

Equation (\ref{72}) already has the expected form of a Heisenberg-Langevin equation for $\hat{a}$ which corresponds to the standard cavity optomechanics Hamiltonian \cite{AKM}
\begin{eqnarray}
H_{\mathrm{om}}&=&\hbar\omega'_c\hat{a}^{\dag}\hat{a}+\hbar g\left(\hat{b}+\hat{b}^{\dag}\right)\hat{a}^{\dag}\hat{a}+\hbar\omega_{\mathrm{m}}\hat{b}^{\dag}\hat{b}
\nonumber\\
&+&\hbar\left(\Omega e^{-i\omega_L t}\hat{a}^{\dag}+\mathrm{h.c.}\right),
\label{H1}
\end{eqnarray}
with $\omega'_c=\omega_c+\Delta_{\mathrm{AC}}$ and $g$ from Eq. (\ref{71}). This Hamiltonian describes a cavity whose resonant frequency is shifted by the motion of an internal membrane reflector (Fig. 4a), leading to a linear (first-order) optomechanical coupling to the corresponding mechanical mode $\hat{b}$.

\begin{figure}[t]
  \begin{center}
    \includegraphics[width=\columnwidth]{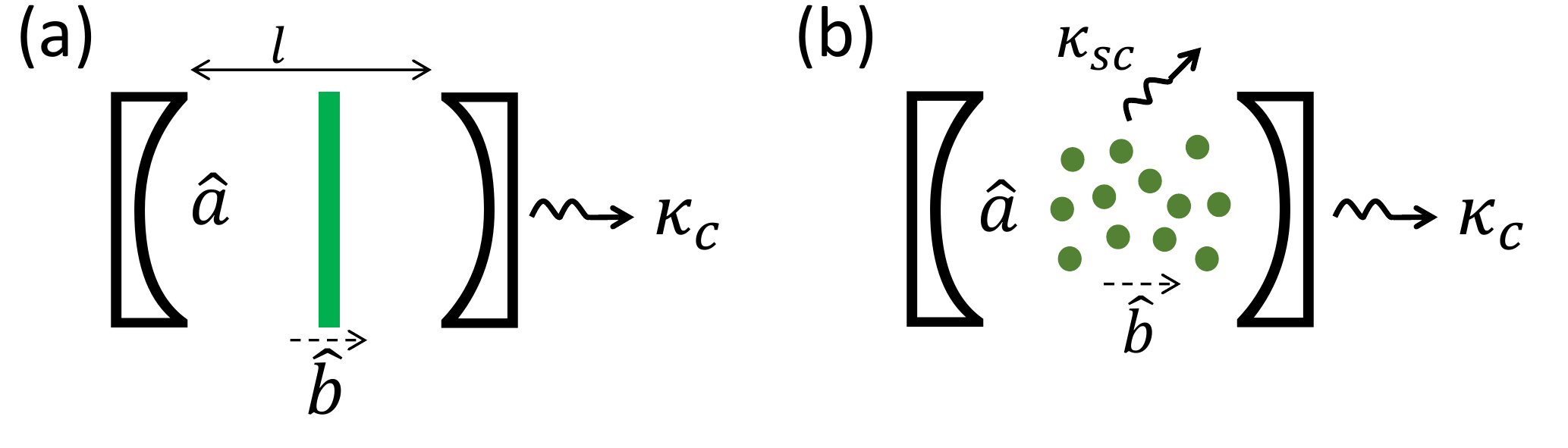}
    \caption{\small{
Cavity optomechanics. (a) Typical cavity optomechanics setup with an intra-cavity solid membrane (reflector) whose motion (bosonic mode $\hat{b}$) modifies the cavity resonance, as captured by the standard Hamiltonian (\ref{H1}) \cite{AKM,HAR}. (b) Same, with an atomic cloud taking the role of the membrane, where $\hat{b}$ is now associated with the center-of-mass coordinate. Scattering from the disordered cloud to outside, non-confined modes leads to an additional cavity damping $\kappa_{sc}$ (even without atomic motion, Fig. 2a).
    }} \label{fig4}
  \end{center}
\end{figure}

The usual damping out of the cavity mirrors, described by $\kappa_c$ and $\hat{F}_c(t)$, is supplemented by damping due to the scattering of photons to the non-confined modes. At zeroth-order in the motion this latter damping vanishes due to the spatial order of the array (as explained in Sec. IV). However, the motion $\hat{z}_n$ of individual atoms gives rise to weak disorder of the perfect ordered array, and hence to residual scattering at second order in motion. The coefficient $\hat{K}_{sc}$, which depends on $q\hat{z}_n$ to second order, contains the information on this additional cavity damping, whereas $\hat{F}_{sc}(t)$ is a corresponding quantum Langevin noise. The expression for $\hat{K}_{sc}$ is found in Appendix B and can used to obtain an estimate of the cavity damping rate to non-confined modes,
\begin{equation}
\kappa_{sc}=\mathrm{Re}[\langle\hat{K}_{sc}\rangle].
\label{ksc1}
\end{equation}
Assuming small enough atomic motion (weak drive $\Omega$), the averaging is performed using the ground state $|0\rangle$ of the motion of all atoms inside the traps, $\hat{b}_n |0\rangle=0$, with $\langle \hat{z}_n\rangle=0$ and $\langle \hat{z}_n\hat{z}_m\rangle=\delta_{nm}x_0^2$. This calculation is described in Appendix C 1, yielding,
\begin{equation}
\kappa_{sc}=\eta^2N_a\frac{c}{l}\left(\frac{\gamma}{\delta-\Delta}\right)^2\frac{\varepsilon/2}{q^2w^2},
%\quad \varepsilon=6\left[\cos^2(qz_0)+\frac{2}{5}\sin^2(qz_0)\right],
\label{ksc}
\end{equation}
with $\varepsilon=6[\cos^2(qz_0)+\frac{2}{5}\sin^2(qz_0)]$ and with $N_a$ from Eq. (\ref{eta}). This expression reflects that motion-induced disorder of the array leads to damping of the cavity, but only to second order in the motion, $\kappa_{sc}\propto \eta^2$.

Finally, the corresponding Langevin noise due to this motion-induced scattering is given by
\begin{eqnarray}
&&\hat{F}_{sc}(t)=-i2\sum_n g^{\ast}_n
\nonumber\\
&&\times\left[\sin(qz_0)\left(1-\frac{q^2\hat{z}_n^2}{2}\right)+\cos(q z_0)q\hat{z}_n\right]\widehat{\delta\sigma}_n(t),
\label{59}
\end{eqnarray}
with $\widehat{\delta\sigma}_n(t)=(1/\sqrt{N})\sum_n e^{-i\mathbf{k}_{\bot}\cdot\mathbf{r}^{\bot}_n}\widehat{\delta\sigma}_{\mathbf{k}_{\bot}}(t)$ and $\widehat{\delta\sigma}_{\mathbf{k}_{\bot}}(t)$ from Eq. (\ref{53a}) [and (\ref{54})]. By consistency of the fluctuation dissipation theorem, it should satisfy
\begin{equation}
\langle \hat{F}_{sc}(t)\hat{F}_{sc}^{\dag}(t')\rangle\approx \kappa_{sc}\delta(t-t').
\label{Fsc}
\end{equation}
Since $\kappa_{sc}\propto\eta^2$ this means that $\hat{F}_{sc}$ does not contain a contribution at zeroth-order in $\hat{z}_n$, as we indeed verify, using Eq. (\ref{30a}).

\emph{Conclusion.---} So far we have derived the equation of motion (\ref{72}) for the cavity mode, which is consistent with the standard optomechanical model (\ref{H1}), and with the optomechanical parameters $g\propto\eta$ from (\ref{71}) and a generally weaker, motion-induced damping $\kappa_{sc}\propto \eta^2$. A more thorough and consistent approach however requires to also find the corresponding dynamical equation for the motion variable $\hat{b}$ and to show that it can be derived from the same Hamiltonian. Such a complete treatment of the motion and the cavity is presented in the following sections.

\section{Multimode cavity optomechanics}
In this section, we extend the description of optomechanics to include the dynamical equations for the motion of the atoms. We show that the resulting coupled equations for the cavity mode and the atomic motion are consistent with a cavity optomechanics model with multiple mechanical degrees of freedom.

\subsection{Equations for atomic motion $\hat{b}_n$}
We begin by writing the dynamical equations for atomic motion, $\hat{\hat{p}}_n$ from Eq. (\ref{27}) together with $\dot{\hat{z}}_n=\hat{p}_n/M$, in terms of the bosonic operators $\hat{b}_n$ from Eq. (\ref{bn}). We then insert the solution for $\tilde{\sigma}_n$ from Eq. (\ref{53a}) [with the transformation (\ref{sigk}) and with (\ref{53b})] into the equation for $\hat{b}_n$, keeping terms up to first order in $q\hat{z}_n$ as explained above and simplifying expressions as in Sec. VI B. Moreover:
\\
(i) Terms which are quadratic in the vacuum-induced noise, $\sim \hat{F}_n^{\dag}\hat{F}_n$, are neglected, since they are related to forces purely induced by the vacuum.
\\
(ii) Terms of order $\sim \hat{a}^{\dag}\hat{F}_n$ are neglected assuming the incident field is weak, $\Omega\ll \kappa_c$, so that the occupation of the cavity is very low $\langle \hat{a}\rangle \ll 1$. For larger driving, where the average cavity field $\langle \hat{a}\rangle$ is significant, terms such as $\sim \hat{a}^{\dag}\hat{F}_n$ lead to a fluctuating force and correspondingly to friction of the motion which scales as $|\langle \hat{a}\rangle|^2$. So, neglecting these terms essentially means that we assume that the average cavity field is weak enough to neglect this friction and associated fluctuations.

The resulting equation for $\hat{b}_n$ becomes
\begin{eqnarray}
\dot{\hat{b}}_n&=&-i\omega_{\mathrm{m}}\hat{b}_n-i\eta \bar{g}\hat{a}^{\dag}\hat{a}V^0_n\left[\sin(2qz_0)-2\sin^2(qz_0)q\hat{z}_n\right]
\nonumber\\
&+&i\eta\bar{g}\hat{a}^{\dag}\hat{a}\sum_m\mathcal{M}_{nm}\sqrt{V^0_nV^0_m}q\hat{z}_m,
\label{88}
\end{eqnarray}
where the effective cavity-mediated, inter-atom mechanical coupling $\mathcal{M}_{nm}$ is given in Appendix B.

\subsection{Collective mechanical modes}
In Eq. (\ref{72}), we identify the relevant mechanical mode $\hat{b}$ that couples to the cavity field. This motivates to define a new set of orthonormal mechanical modes $\{\nu\}$ with spatial profiles $V_n^{\nu}$ (real) spanning the space $\{n\}$, which includes the relevant mode $\hat{b}=\hat{b}_{\nu=0}$ with the profile $V_n^0$ from Eq. (\ref{73}),
\begin{eqnarray}
&&\hat{b}_{\nu}=\sum_n V_n^{\nu}\hat{b}_n, \quad \quad \sum_n V_n^{\nu}V_n^{\nu'}=\delta_{\nu\nu'},
\nonumber\\
&&\hat{b}_n=\sum_{\nu}V_n^{\nu}\hat{b}_{\nu}, \quad\quad \sum_{\nu} V_n^{\nu}V_m^{\nu}=\delta_{nm}.
\nonumber\\
\label{bj}
\end{eqnarray}
Equation (\ref{88}) is then transformed to a dynamical equation for $\hat{b}_{\nu}$. For the corresponding equation for $\tilde{a}$ in (\ref{72}), while considering the explicit expression for $\hat{K}_{sc}$ [Eq. (\ref{70}) in Appendix B], we express all terms that contain $\hat{z}_n=x_0(\hat{b}_n+\hat{b}_n^{\dag})$ using the collective modes $\hat{b}_{\nu}$. We then obtain the coupled dynamical equations for the cavity mode and the collective mechanical modes as
\begin{eqnarray}
\dot{\tilde{a}}&=&\left[i(\delta_c-\Delta_{\mathrm{AC}})-\frac{\kappa_c}{2}\right]\tilde{a}-ig\left(\hat{b}+\hat{b}^{\dag}\right)\tilde{a}
-i\Omega+\hat{F}_c(t)
\nonumber\\
&+&\sum_{\nu\nu'}C_{\nu\nu'}\left(\hat{b}_{\nu}+\hat{b}^{\dag}_{\nu}\right)\left(\hat{b}_{\nu'}+\hat{b}^{\dag}_{\nu'}\right)\tilde{a}
+\hat{F}_{sc},
\nonumber\\
\dot{\hat{b}}_{\nu}&=&-i\omega_{\mathrm{m}}\hat{b}_{\nu}-ig\delta_{\nu0}\tilde{a}^{\dag}\tilde{a}+i2\sum_{\nu'}\mathrm{Im}[C_{\nu\nu'}]\left(\hat{b}_{\nu'}+\hat{b}_{\nu'}^{\dag}\right)\tilde{a}^{\dag}\tilde{a},
\nonumber\\
\label{94}
\end{eqnarray}
with the first-order optomechanical coupling $g\propto \eta$ from Eq. (\ref{71}) and where the expression for the second-order couplings $C_{\nu\nu'}\propto \eta^2$ is given in Eq. (\ref{95}) of Appendix B.

\subsection{Multimode cavity optomechanics}
The coupled equations (\ref{94}) represent an optomechanical interaction up to second order between the collective mechanical modes $\hat{b}_{\nu}$ and the cavity mode $\hat{a}$. Importantly, the conservative optomechanical part of both equations can be derived from a single Hamiltonian,
\begin{eqnarray}
H_{\mathrm{eff}}&=&\hbar g\hat{a}^{\dag}\hat{a}\left(\hat{b}+\hat{b}^{\dag}\right)
\nonumber\\
&-&\hbar\sum_{\nu\nu'}\mathrm{Im}[C_{\nu\nu'}]\hat{a}^{\dag}\hat{a}
\left(\hat{b}_{\nu}+\hat{b}^{\dag}_{\nu}\right)\left(\hat{b}_{\nu'}+\hat{b}^{\dag}_{\nu'}\right).
\label{Hmm}
\end{eqnarray}
In addition, the equation for $\hat{a}$ in (\ref{94}) contains an optomechanically induced non-conservative part, $\propto \mathrm{Re}[C_{\nu\nu'}]$, which cannot be derived from the Hamiltonian (\ref{Hmm}), and which is shown below to result in the additional cavity damping $\kappa_{sc}$ related to the Langevin noise $\hat{F}_{sc}$.

\emph{Light-induced dipole-dipole forces.---} The second term in the Hamiltonian (\ref{Hmm}) describes the mechanical coupling between the different collective coordinates $\hat{b}_{\nu}$, with a coupling ``spring constant" $2\mathrm{Im}[C_{\nu\nu'}]\hat{a}^{\dag}\hat{a}$, proportional to the intensity of the light $\hat{a}^{\dag}\hat{a}$ and to second-order derivatives of the dipole-dipole interaction. Hence, this mechanical coupling is interpreted as light-induced dipole-dipole forces \cite{THI,SAL,LIDDI}, which were recently shown to lead to multimode optomechanics for an atom array even without a cavity \cite{om}.

\section{Mapping to standard cavity optomechanics}
Within the multimode optomechanical description of Eqs. (\ref{94}) and (\ref{Hmm}), only the collective mechanical mode $\hat{b}$ ($\nu=0$) is coupled to the cavity at linear order, whereas all the rest of the modes have a weaker quadratic coupling. This reflects the intuition that only the mode $\hat{b}$, which spatially follows the intensity profile of the cavity mode, is directly coupled to the cavity, and therefore motivates to move to a single-mechanical-mode picture as in the standard optomechanics Hamiltonian (\ref{H1}).

\subsection{Single mechanical mode}
Since the optomechanical coupling of the single mechanical mode $\hat{b}$, $g\propto \eta$, is much stronger than those of the rest of the mechanical modes, $C_{\nu\nu'}\propto \eta^2$, we can eliminate the latter as weakly coupled reservoir modes.
To that end, we first formally solve Eq. (\ref{94}) for $\hat{b}_{\nu \neq 0}$,
\begin{eqnarray}
\hat{b}_{\nu}(t)=\hat{b}_{\nu}(0)e^{-i\omega_{\mathrm{m}} t}+\mathcal{O}(\eta^2),
\label{97}
\end{eqnarray}
and insert this solution into Eq. (\ref{94}) for $\hat{b}$ ($\nu=0$),
\begin{eqnarray}
\dot{\hat{b}}&=&-i\omega_{\mathrm{m}}\hat{b}-i\left[g-2\mathrm{Im}[C_{00}]\left(\hat{b}+\hat{b}^{\dag}\right)\right]\tilde{a}^{\dag}\tilde{a}
\nonumber\\
&&+\hat{f}(t)+\mathcal{O}(\eta^4),
\label{98}
\end{eqnarray}
where
\begin{eqnarray}
\hat{f}(t)=i2\sum_{\nu\neq0}\mathrm{Im}[C_{0\nu}]\left(\hat{b}_{\nu}(0)e^{-i\omega_{\mathrm{m}} t}+\hat{b}^{\dag}_{\nu}(0)e^{i\omega_{\mathrm{m}} t}\right)\tilde{a}^{\dag}\tilde{a}
\nonumber\\
\label{118}
\end{eqnarray}
is a force originating from the light-induced dipole-dipole force between the coordinate $\hat{b}$ and all the rest of the coordinates $\nu\neq 0$, and therefore depends on the quantum state of the modes. Assuming that initially all of the atoms are in their mechanical ground state, $|0\rangle$, i.e. $\hat{b}_n(0)|0\rangle=0$, we obtain the statistics,
$\langle \hat{b}_{\nu}(0)\rangle,\langle \hat{b}_{\nu}(0)\hat{b}_{\nu'}(0)\rangle,\langle \hat{b}^{\dag}(0)\hat{b}_{\nu'}(0)\rangle=0$ and
$\langle \hat{b}_{\nu}(0)\hat{b}^{\dag}_{\nu'}(0)\rangle=\delta_{\nu\nu'}$.
The force $\hat{f}(t)$ from Eq. (\ref{118}) is then a zero-mean Langevin force with correlation $\langle \hat{f}(t)\hat{f}(t')\rangle = \mathcal{O}(\eta^4)$. By the fluctuation-dissipation theorem, the fluctuating force $\hat{f}(t)$ is associated with the friction term of order $\mathcal{O}(\eta^4)$ that appears unspecified in Eq. (\ref{98}). Since we wish to keep terms only up to second order in the motion $\eta^2$, this friction and the associated Langevin force are then neglected and we have
\begin{eqnarray}
\dot{\hat{b}}\approx -i\omega_{\mathrm{m}}\hat{b}-ig\tilde{a}^{\dag}\tilde{a}
-i 2g_2\left(\hat{b}+\hat{b}^{\dag}\right)\tilde{a}^{\dag}\tilde{a},
\label{103}
\end{eqnarray}
with the second-order optomechanical coupling $g_2=-\mathrm{Im}[C_{00}]$, which is found to be (see Appendix C 3)
\begin{eqnarray}
g_2=\eta^2\frac{c}{l}\frac{\gamma}{\delta-\Delta}\frac{4}{q^2w^2}\cos^2(qz_0).
\label{g2}
\end{eqnarray}

Turning to the equation for the cavity mode, we insert the solution for $\hat{b}_{\nu \neq 0}$ from Eq. (\ref{97}) into Eq. (\ref{94}) for $\tilde{a}$ and keep terms to second order in $\eta^2$ [amounting to the replacement $\hat{b}_{\nu \neq 0}(t)\rightarrow \hat{b}_{\nu}(0)e^{-i\omega_{\mathrm{m}} t}$]. We then average over these mechanical ``reservoir" degrees of freedom, obtaining
\begin{eqnarray}
\dot{\tilde{a}}&=&\left[i\left(\delta_c-\Delta_{\mathrm{AC}}-\Delta_{sc}\right)-\frac{\kappa_{c}+\kappa_{sc}}{2}\right]\tilde{a}
-ig\left(\hat{b}+\hat{b}^{\dag}\right)\tilde{a}
\nonumber\\
&-&ig_2\tilde{a}\left(\hat{b}+\hat{b}^{\dag}\right)^2 \tilde{a}
-i\Omega+\hat{F}_c+\hat{F}_{sc},
\label{111}
\end{eqnarray}
with
\begin{eqnarray}
\kappa_{sc}=2\sum_{\nu\neq0}\mathrm{Re}[C_{\nu\nu}]\propto \eta^2, \:\,\, \Delta_{sc}=\sum_{\nu\neq0}\mathrm{Im}[C_{\nu\nu}]\propto \eta^2.
\label{112}
\end{eqnarray}
Importantly, in Appendix C 2 we show that $\kappa_{sc}=2\sum_{\nu\neq0}\mathrm{Re}[C_{\nu\nu}]$ here gives the same expression for $\kappa_{sc}$ found in Eq. (\ref{ksc}). Here, however, we find $\kappa_{sc}$ from a consistent description of the mechanical degrees of freedom.
Finally, the second-order frequency shift $\Delta_{sc}\sim \eta^2\Delta_{\mathrm{AC}}\ll \Delta_{\mathrm{AC}}$ is neglected in the following.

\subsection{Effective cavity optomechanics model}
The above considerations lead to standard-form coupled dynamical equations for the cavity mode $\hat{a}$ and collective atomic coordinate $\hat{b}$, up to second order in the motion,
\begin{eqnarray}
\dot{\hat{a}}&=&-\left(i\omega'_c-\frac{\kappa}{2}\right)\hat{a}-ig\left(\hat{b}+\hat{b}^{\dag}\right)\hat{a}
\nonumber\\
&-&ig_2\hat{a}\left(\hat{b}+\hat{b}^{\dag}\right)^2
+\left[-i\Omega+\hat{F}(t)\right]e^{-i\omega_L t},
\nonumber\\
\dot{\hat{b}}&=&-i\omega_{\mathrm{m}}\hat{b}-ig\hat{a}^{\dag}\hat{a}-i2g_2\left(\hat{b}+\hat{b}^{\dag}\right)^2\hat{a}^{\dag}\hat{a},
\label{115}
\end{eqnarray}
with (neglecting $\Delta_{sc}$)
\begin{eqnarray}
\omega'_c=\omega_c+\Delta_{\mathrm{AC}}, \:\,\, \kappa=\kappa_c+\kappa_{sc}, \:\,\, \hat{F}(t)=\hat{F}_c(t)+\hat{F}_{sc}(t),
\nonumber\\
\label{par}
\end{eqnarray}
and where $\kappa_{sc}$, $g$ and $g_2$ are those from Eqs. (\ref{ksc}), (\ref{71}) and (\ref{g2}), respectively.
Using Eqs. (\ref{Fc}) and (\ref{Fsc}) and the fact that the Langevin noises $\hat{F}_c$ and $\hat{F}_{sc}$ are uncorrelated (originate from orthogonal photon modes), we have
\begin{eqnarray}
\langle\hat{F}(t)\hat{F}^{\dag}(t')\rangle=\kappa \delta(t-t'),
\label{FFF}
\end{eqnarray}
consistent with the fluctuation-dissipation theorem.

Moreover, it is easy to verify that the conservative part of Eqs. (\ref{115}) can be derived from the standard cavity optomechanical Hamiltonian as in (\ref{H1}), also including second-order optomechanical coupling,
\begin{eqnarray}
H'_{\mathrm{om}}&=&\hbar\omega'_c\hat{a}^{\dag}\hat{a}+\hbar\omega_{\mathrm{m}}\hat{b}^{\dag}\hat{b}
+\hbar\left(\Omega e^{-i\omega_L t}\hat{a}^{\dag}+\mathrm{h.c.}\right)
\nonumber\\
&+&\hbar g\left(\hat{b}+\hat{b}^{\dag}\right)\hat{a}^{\dag}\hat{a}
+\hbar g_2\left(\hat{b}+\hat{b}^{\dag}\right)^2\hat{a}^{\dag}\hat{a}.
\label{HH}
\end{eqnarray}

\subsection{Cavity optomechanics of an atom-array membrane}
Starting from a many-atom system of an array inside a cavity, and by eliminating the detuned atomic internal states, we were able to identify a single relevant mechanical mode and consistently eliminate all others, arriving at the standard cavity optomechanics description of Hamiltonian (\ref{HH}). This establishes an array of atoms inside a cavity as a novel realization of the membrane-in-the-middle setup of optomechanics \cite{HAR}, where the role of the membrane is played by the 2D atom array. The potential advantages and novel opportunities opened by this system are discussed in our work, Ref. \cite{omc}. In a nutshell, the main advantage of this system lies in its combination of the large mechanical susceptibility of trapped atoms with the spatial order of an array. This is illustrated by a comparison with the two other realizations of cavity optomechanics shown in Fig. 4. The membrane from Fig. 4a is made of a bulk, solid material with a very small zero-point motion $x_0$ and hence small optomechanical coupling $g\propto\eta \propto x_0$. In contrast, the atoms of the array are trapped by lasers and have much larger $x_0$ and $g$. This is also the case for the atom cloud in Fig. 4b. However, for a disordered cloud, the large $g$ comes at the price of increased cavity losses due to scattering from the atoms, $\kappa_{sc}$. For the atom array, this scattering arises only due to motion and scales as $\eta^2$, becoming much smaller than the desired optomechanical coupling $g\propto\eta$. In Ref. \cite{omc} we show that this favorable scaling can lead to observable quantum optomechanical effects at the single-photon level.

\section{Discussion}
To conclude, we developed a cavity QED formalism wherein cooperative effects due to the interaction of an intra-cavity atomic array with the outside, non-transversely-confined electromagnetic modes are taken into account. These cooperative effects, which are typically ignored, are crucial for an ordered array of emitters. We began building our formalism considering the ``ideal" case of linearly-responding atoms in a perfectly-ordered, motionless array, which lead us to the basic conditions (\ref{30}).
For this ``ideal" case, scattering to outside, non-confined modes is totally suppressed. However, the condition (\ref{30}) is useful also beyond the \emph{motionless} and \emph{linear} case. As a first step in that direction, we considered the \emph{motion} of atoms, which causes an effective disorder in the array structure. The condition (\ref{30}) and its extension to small-amplitude motion (\ref{31}) were then used to analyze this case, finding novel opportunities in optomechanics.

A second, analogous step beyond the ``ideal" case will be to consider the \emph{nonlinearity} of the internal atomic response. This also introduces an effective disorder since excited (saturated) atoms exhibit a different polarizability than ground-state atoms. For weak enough incident light and lowest-order nonlinearity, we would expect that condition (\ref{30}) still plays an important role, possibly with an extension in analogy to (\ref{31}).

Finally, it should be noted that our approach is applicable to treating 2D atom arrays in other transversely confined geometries, beyond the cavity case. Considering general situations where the set of electromagnetic modes of interest are transversely confined and where the rest of the non-transversely-confined modes are treated as a reservoir, an equation similar to (\ref{30}) should hold, as long as the system geometry that defines the confined modes does not mix them with the non-confined modes. Examples may include a single mirror, an array of mirrors, or even a waveguide structure.

\begin{acknowledgments}
We acknowledge fruitful discussions with Peter Rabl, and financial support from the NSF, the MIT-Harvard Center for Ultracold Atoms, the Vannevar Bush Faculty Fellowship, and a research grant from the
Center for New Scientists at the Weizmann Institute of Science.
\end{acknowledgments}

\appendix
\section{Derivatives of the approximated Green's function}
The expression for the Green's function in (\ref{31a}) contains the free-space and the transversely confined parts as follows.
\subsection{Free-space Green's function}
The free-space part is given by
\begin{equation}
D^{\mathrm{fs}}(\mathbf{r}-\mathbf{r}')=-i\frac{3}{2}\gamma\lambda \mathbf{e}_d^{\dag}\cdot \overline{\overline{G}}_{\mathrm{fs}}(\mathbf{r}-\mathbf{r}')\cdot\mathbf{e}_d,
\label{A1}
\end{equation}
where $\overline{\overline{G}}_{\mathrm{fs}}$ is the dyadic Green's function tensor \cite{NH}
\begin{eqnarray}
&&\left[\overline{\overline{G}}_{\mathrm{fs}}(\mathbf{r})\right]_{ij}
\nonumber\\
&&=\frac{e^{iqr}}{4\pi r}\left[\left(1+\frac{iqr-1}{q^2r^2}\right)\delta_{ij}+\left(-1+\frac{3-3iqr}{q^2r^2}\right)\frac{r^ir^j}{r^2}\right],
\nonumber\\
\label{Gfs}
\end{eqnarray}
with $i,j\in\{x,y,z\}$, $r=|\mathbf{r}|$ and $r^i=\mathbf{e}_i\cdot\mathbf{r}$. Considering the expansion around equilibrium atomic positions, with $z_n=z_m=z_0$ (i.e. around $z=z_n-z_m=0$), the first derivative with respect to $z$ vanishes, whereas the second derivative is given by
\begin{equation}
D_{\mathrm{fs}}''(\mathbf{r})\equiv\left.\frac{\partial^2}{\partial z^2}D^{\mathrm{fs}}(\mathbf{r})\right|_{z=0}
=-i\frac{3}{4}q^2\gamma \mathbf{e}_d^{\dag} \cdot \overline{\overline{F}}(\mathbf{r}_{\bot})\cdot \mathbf{e}_d,
\label{Fnm}
\end{equation}
where $\overline{\overline{F}}(\mathbf{r}_{\bot})$ is a dipole-dipole tensor given in Eq. (A4) of Ref. \cite{AAMO}. For the calculations in Appendix C below, we need to evaluate  $\mathrm{Re}[D_{\mathrm{fs}}''(\mathbf{r}=0)]$. Taking a circularly-polarized dipole-transition, $\mathbf{e}_d=(\mathbf{e}_x+i\mathbf{e}_y)/\sqrt{2}$, we find $\lim_{\mathbf{r}_{\bot}\rightarrow 0} \mathrm{Im}[\mathbf{e}_d^{\dag} \cdot \overline{\overline{F}}(\mathbf{r}_{\bot})\cdot \mathbf{e}_d]=-4/15$, and
\begin{equation}
\mathrm{Re}[D_{\mathrm{fs}}''(\mathbf{r}=0)]=-\frac{1}{5}q^2\gamma.
\label{Dpp0}
\end{equation}

\subsection{Transversely confined Green's function}
This part of the Green's function is defined by
\begin{eqnarray}
D^c(\mathbf{r},\mathbf{r}')&=&-i\frac{3}{2}\gamma\lambda\sum_{\mu\in\mathrm{c}}\frac{u_\mu(\mathbf{r})u^{\ast}_\mu(\mathbf{r}')}{(\omega_\mu/c)^2-q^2}.
\label{Dc}
\end{eqnarray}
The transversely-confined subspace ``c" includes the cavity-confined modes + the transversely matching outside modes that they are coupled to via the mirrors. This subspace can be  spanned by propagating, transversely confined paraxial modes of the form
\begin{eqnarray}
u_{p p' k s}(\mathbf{r})=\varphi_{p p' ks}(\mathbf{r}_{\bot},z)e^{i s k z}.
\label{uc}
\end{eqnarray}
Here the mode indices $\mu \rightarrow p, p',k,s$ include the two transverse-profile indices $pp'$, the mode frequency $k=\omega/c$ and the propagation direction $s=\pm$ (right/left). The envelopes $\varphi_{p p' ks}$ vary slowly with $z$ at a length-scale $z_R=\pi w^2/\lambda\gg \lambda$ around the focus point $z=0$, where $w\gg\lambda$ is a scale of transverse confinement at the focus (the waist).
One concrete example could be the Hermite-Gauss modes, for which
\begin{eqnarray}
\varphi_{p p' k s}(\mathbf{r}_{\bot},z)&\propto&
\frac{w}{\tilde{w}(z)}e^{-\frac{r_{\bot}^2}{\tilde{w}^2(z)}}H_{p}\left(\frac{\sqrt{2}x}{\tilde{w}(z)}\right)H_{p'}\left(\frac{\sqrt{2}y}{\tilde{w}(z)}\right)
\nonumber\\
&& \times
 e^{i s \frac{k r_{\bot}^2}{2 R(z)}}e^{-i s \psi_{p p'}(z)},
\label{HG}
\end{eqnarray}
with $\tilde{w}(z)$ and $R(z)$ from Sec. II B, $\psi_{pp'}(z)=(1+p+p')\arctan(z/z_R)$, and $H_p(x)$ being the Hermite polynomial of order $p$.

Inserting Eqs. (\ref{Dc}) and (\ref{uc}) into the expansion of $D$ and hence $D^c(\mathbf{r},\mathbf{r}')$ around the equilibrium array positions $\mathbf{r}^{(0)}_n=(\mathbf{r}^{\bot}_n,z_0)$, Eq. (\ref{26D}), the differentials of the type
\begin{eqnarray}
\left.\frac{\partial}{\partial z} u_{pp'ks}(\mathbf{r})\right|_{\mathbf{r}^{(0)}_n} z_n
=isk z_n u_{pp'ks}(\mathbf{r}^{(0)}_n)+ \mathcal{O}(z_n/z_R)
\nonumber\\
\label{diff}
\end{eqnarray}
appear. The second term originates from $\left.\partial_z \varphi_{p p' k s}(\mathbf{r}_{\bot},z)\right|_{\mathbf{r}^{(0)}_n}\propto1/z_R$, using the assumption that the equilibrium position of the atoms $z=z_0$ is close enough to the focus ($z_0\ll z_R$), where the envelope varies in a scale $z_R$. Then, since the poles in Eq. (\ref{Dc}) dictate $k\sim q=2\pi/\lambda$, and since $z_R\gg \lambda$, the second term in (\ref{diff}) is negligible, and the expansion (\ref{26D}) written for $D^c$ becomes
\begin{eqnarray}
D^c(\hat{\mathbf{r}}_n,\hat{\mathbf{r}}_m)&\approx& D^c_{nm}+\sum_{pp'ks}
\frac{u_{p p' k s}(\mathbf{r}^{(0)}_n)u^{\ast}_{p p' k s}(\mathbf{r}^{(0)}_m)}{k^2-q^2}
\nonumber\\
&&\times\left[isk(\hat{z}_n-\hat{z}_m)-\frac{k^2}{2}(\hat{z}_n-\hat{z}_m)^2\right].
\nonumber\\
\label{B5}
\end{eqnarray}
Since the envelope $\varphi_{p p' k s}(\mathbf{r}_{\bot},z)$ is independent of $s$ at $z=0$, which is approximately true also for the equilibrium position $z=z_0\ll z_R$, the linear term $\propto(\hat{z}_n-\hat{z}_m)$ is negligibly small. Together with the vanishing linear-order in the expansion for the free-space part of $D$ (Appendix A 1 above), this leads to Eq. (\ref{33c}).

\section{Explicit expressions of some terms and coefficients}
Here we provide the explicit expressions for a few terms and coefficients that appear in Secs. VI and VII.

\subsection{Sec. VI: Elimination of internal states}
The steady-state solution for $\tilde{\sigma}_{\mathbf{k}_{\bot}}$ from Eq. (\ref{53a}) includes the following terms. The coefficient $\hat{W}_{\mathbf{k}_{\bot}}$ is an operator that contains zeroth-, first-, and second-order contributions in the small parameter $q\hat{z}_n$, given by
\begin{eqnarray}
\hat{W}_{\mathbf{k}_{\bot}}&=&
\sin(qz_0)\left[\frac{2g_{\mathbf{k}_{\bot}}}{\delta_{\mathbf{k}_{\bot}}}-i\sum_{\mathbf{k}'_{\bot}}\frac{\gamma_{\mathbf{k}_{\bot}\mathbf{k}'_{\bot}}g_{\mathbf{k}'_{\bot}}}{\delta_{\mathbf{k}_{\bot}}^2}\right]
\nonumber\\
&+&\cos(qz_0)\left[\frac{2}{\delta_{\mathbf{k}_{\bot}}}\frac{1}{\sqrt{N}}\sum_n e^{-i\mathbf{k}_{\bot}\cdot\mathbf{r}^{\bot}_n}g_nq\hat{z}_n
\right.\nonumber\\
&&\left.-i\sum{\mathbf{k}'_{\bot}}\frac{\gamma_{\mathbf{k}_{\bot}\mathbf{k}'_{\bot}}}{\delta_{\mathbf{k}_{\bot}}^2}\frac{1}{\sqrt{N}}\sum_n e^{-i\mathbf{k}'_{\bot}\cdot\mathbf{r}^{\bot}_n}g_n q\hat{z}_n\right]
\nonumber\\
&-&\sin(qz_0)\left[\frac{1}{\delta_{\mathbf{k}_{\bot}}}\frac{1}{\sqrt{N}}\sum_n e^{-i\mathbf{k}_{\bot}\cdot\mathbf{r}^{\bot}_n}g_nq^2\hat{z}^2_n
\right.\nonumber\\
&&\left.-\frac{i}{2}\sum{\mathbf{k}'_{\bot}}\frac{\gamma_{\mathbf{k}_{\bot}\mathbf{k}'_{\bot}}}{\delta_{\mathbf{k}_{\bot}}^2}\frac{1}{\sqrt{N}}\sum_n e^{-i\mathbf{k}'_{\bot}\cdot\mathbf{r}^{\bot}_n}g_n q^2\hat{z}^2_n
\right.\nonumber\\
&&\left.+i2\sum{\mathbf{k}'_{\bot}}\frac{g_{\mathbf{k}'_{\bot}}}{\delta_{\mathbf{k}_{\bot}}\delta_{\mathbf{k}'_{\bot}}}\hat{J}_{\mathbf{k}_{\bot}\mathbf{k}'_{\bot}}\right],
\label{53b}
\end{eqnarray}
with $\delta_{\mathbf{k}_{\bot}}\equiv \delta-\Delta_{\mathbf{k}_{\bot}}$ and $g_{\mathbf{k}_{\bot}}=(1/\sqrt{N})\sum_ne^{-i\mathbf{k}_{\bot}\cdot\mathbf{r}^{\bot}_n}g_n$, and recalling that $\hat{J}_{\mathbf{k}_{\bot}\mathbf{k}'_{\bot}}\propto\hat{J}_{nm}\propto (\hat{z}_n-\hat{z}_m)^2$ [Eq. (\ref{33c})].

The noise term $\widehat{\delta\sigma}_{\mathbf{k}_{\bot}}$ from Eq. (\ref{53a}) originates from the Langevin operator due to the non-confined modes  $\hat{F}_{\mathbf{k}_{\bot}}$, and is given by
\begin{eqnarray}
\widehat{\delta\sigma}_{\mathbf{k}_{\bot}}&=&\int_0^te^{-i\delta_{\mathbf{k}_{\bot}}(t-t')}\left[
-\sum_{\mathbf{k}'_{\bot}}\frac{\gamma_{\mathbf{k}_{\bot}\mathbf{k}'_{\bot}}+2\hat{J}_{\mathbf{k}_{\bot}\mathbf{k}'_{\bot}}(t)}{2}
\right.\nonumber\\
&&\left.\times\int_0^{t'}dt''e^{-i\delta_{\mathbf{k}'_{\bot}}(t'-t'')}\hat{F}_{\mathbf{k}'_{\bot}}(t'')+\hat{F}_{\mathbf{k}_{\bot}}(t')\right].
\nonumber\\
\label{54}
\end{eqnarray}

Then, to obtain Eq. (\ref{72}) for the cavity mode, we insert Eq. (\ref{53a}), with $\hat{W}_{\mathbf{k}_{\bot}}$ from Eq. (\ref{53b}), into Eq. (\ref{27}) for $\tilde{a}$. For simplifying the different terms, we use the following:
\\
(i) Since the cavity mode is paraxial ($w\gg \lambda$), $g_{\mathbf{k}_\bot}$ is evaluated only over a narrow range around $\mathbf{k}_{\bot}=0$ and we use approximations such as $g_{\mathbf{k}_{\bot}}/(\delta-\Delta_{\mathbf{k}_{\bot}})\approx g_{\mathbf{k}_{\bot}}/(\delta-\Delta)$, with  $\Delta=\Delta_{\mathbf{k}_{\bot}=0}$. Moreover, since $w\gg \lambda>a$ we can convert sums into integrals, e.g. $\sum_n |g_n|^2\approx (1/a^2)\int d\mathbf{r}_{\bot}|g(\mathbf{r}_{\bot})|^2=(\gamma+\Gamma)(c/l)/4$.
\\
(ii) From the condition (\ref{30a}) it is easy to show that $\sum_n\sum_m g^{\ast}_n\gamma_{nm}g_m K_m=0$ for any function of $m$ $K_m$ (noting that $g_n\propto u_{1,n}$). This property is used here for $K_m=1,\hat{z}_m,\hat{z}_m^2$.

The expression for the effective damping operator $\hat{K}_{sc}$, that appears in the resulting equation for (\ref{72}) for the cavity mode, is given by
\begin{eqnarray}
\hat{K}_{sc}&=&-i2\sin^2(q z_0) \bar{g}q^2\sum_n V_n^0\hat{z}_n^2
\nonumber\\
&+&\bar{g}\sum_{n,m}\sqrt{V_n^0 V_m^0}\left[\sin^2(qz_0)\frac{D''_{nm}}{\delta-\Delta}(\hat{z}_n-\hat{z}_m)^2
\right. \nonumber\\
&&\left.+\cos^2(qz_0)q^2\hat{z}_n\hat{z}_m\left(\frac{1}{N}\sum_{\mathbf{k}_{\bot}}
e^{i\mathbf{k}_{\bot}\cdot(\mathbf{r}_n^{\bot}-\mathbf{r}_m^{\bot})}\frac{i2(\delta-\Delta)}{\delta-\Delta_{\mathbf{k}_{\bot}}}
\right. \right.\nonumber\\
&&\left.\left.+\frac{1}{N}\sum_{\mathbf{k}_{\bot}}\sum_{\mathbf{k}'_{\bot}}
e^{i\mathbf{k}_{\bot}\cdot\mathbf{r}_n^{\bot}}e^{-i\mathbf{k}'_{\bot}\cdot\mathbf{r}_m^{\bot}}
\frac{\gamma_{\mathbf{k}_{\bot}\mathbf{k}'_{\bot}}(\delta-\Delta)}{(\delta-\Delta_{\mathbf{k}_{\bot}})^2}\right)\right].
\nonumber\\
\label{70}
\end{eqnarray}

\subsection{Sec. VII: Multimode optomechanics}
Equation (\ref{88}) reveals the coupling between the mechanical degrees of freedom of a pair of atoms $n$ and $m$, with the coupling constant
\begin{eqnarray}
\mathcal{M}_{nm}&=&\sin^2(qz_0)\frac{2\mathrm{Im}[D''_{nm}]}{q^2(\delta-\Delta)}
 \nonumber\\
&&-\cos^2(qz_0)\frac{1}{N}\sum_{\mathbf{k}_{\bot}}\left[e^{-i\mathbf{k}_{\bot}\cdot(\mathbf{r}^{\bot}_n-\mathbf{r}^{\bot}_m)}\frac{\delta-\Delta}{\delta-\Delta_{\mathbf{k}_{\bot}}}
\right. \nonumber\\
&&\left.+\frac{i}{2}\sum_{\mathbf{k}'_{\bot}}e^{-i\mathbf{k}_{\bot}\cdot\mathbf{r}^{\bot}_n}e^{i\mathbf{k}'_{\bot}\cdot\mathbf{r}^{\bot}_m}
\frac{\gamma_{\mathbf{k}_{\bot}\mathbf{k}'_{\bot}}(\delta-\Delta)}{\delta-\Delta_{\mathbf{k}_{\bot}}^2}+\mathrm{h.c.}\right].
\nonumber\\
\label{Mnm}
\end{eqnarray}

Within the description of collective mechanical modes $\hat{b}_{\nu}$, the analogous inter-mode couplings $C_{\nu\nu'}$ from Eq. (\ref{94}) are given by
\begin{eqnarray}
C_{\nu\nu'}&=&\eta^2\bar{g}\left[i\sin^2(qz_0)\sum_nV_n^0 V_n^{\nu} V_n^{\nu'}
\right.\nonumber\\
&&\left.+\sin^2(qz_0)\sum_{nm}\sqrt{V_n^0V_m^0}V_n^{\nu} V_m^{\nu'}\frac{D''_{nm}}{q^2(\delta-\Delta)}
\right.\nonumber\\
&&\left.-i\cos^2(qz_0)\sum_{nm}\sqrt{V_n^0V_m^0}V_n^{\nu} V_m^{\nu'}
\right.\nonumber\\
&&\left.\times\frac{1}{N}\sum_{\mathbf{k}_{\bot}}\left(
e^{i\mathbf{k}_{\bot}\cdot(\mathbf{r}^{\bot}_n-\mathbf{r}^{\bot}_m)}\frac{\delta-\Delta}{\delta-\Delta_{\mathbf{k}_{\bot}}}
\right.\right.\nonumber\\
&&\left.\left.-\frac{i}{2}\sum_{\mathbf{k}'_{\bot}}e^{i\mathbf{k}_{\bot}\cdot\mathbf{r}^{\bot}_n}e^{-i\mathbf{k}'_{\bot}\cdot\mathbf{r}^{\bot}_m}
\frac{\gamma_{\mathbf{k}_{\bot}\mathbf{k}'_{\bot}}(\delta-\Delta)}{\delta-\Delta_{\mathbf{k}_{\bot}}^2}\right)\right],
\nonumber\\
\label{95}
\end{eqnarray}
with $\bar{g}$ from Eq. (\ref{71}).

\section{Second-order optomechanical effects}
Here we discuss the estimation of two second-order effects, the motion-induced decay $\kappa_{sc}$ and the optomechanical coupling $g_2$.

\subsection{Motion-induced decay: Estimation of $\kappa_{sc}$ directly from Eq. (\ref{ksc1})}
Using the definition $\kappa_{sc}=\mathrm{Re}[\langle\hat{K}_{sc}\rangle]$ from Eq. (\ref{ksc1}) and performing the averaging over $\hat{K}_{sc}$ in (\ref{70}) with the mechanical ground state $\langle \hat{z}_n\hat{z}_m\rangle=\delta_{nm}x_0^2$, we have
\begin{eqnarray}
\kappa_{sc}&=&
2\sin^2(qz_0)\bar{g}\eta^2\left[\sum_n V_n^0\frac{-\mathrm{Re}[D''_{nn}]}{q^2(\delta-\Delta)}
\right.\nonumber\\
&&\left.+\sum_{n,m}\sqrt{V_n^0 V_m^0}\frac{\mathrm{Re}[D''_{nm}]}{q^2(\delta-\Delta)}\right]
\nonumber\\
&+&\cos^2(qz_0)\bar{g}\eta^2\sum_nV_n^0
\frac{1}{N}
\nonumber\\
&&\times\mathrm{Re}\left[\sum_{\mathbf{k}_{\bot}}\sum_{\mathbf{k}'_{\bot}}e^{i(\mathbf{k}_{\bot}-\mathbf{k}'_{\bot})\cdot\mathbf{r}_n^{\bot}}
\frac{\gamma_{\mathbf{k}_{\bot}\mathbf{k}'_{\bot}}(\delta-\Delta)}{(\delta-\Delta_{\mathbf{k}_{\bot}})^2}\right].
\nonumber\\
\label{C3}
\end{eqnarray}
For the first term in (\ref{C3}) we need to estimate $\mathrm{Re}[D''_{nn}]$, which represents the second derivative of the individual-atom emission. Since this individual-atom term is not expected to vanish in general, we can resort to the typical approximation and evaluate it using the free space result from Eq. (\ref{Dpp0}), i.e. $\mathrm{Re}[D''_{nn}]\approx \mathrm{Re}[D_{\mathrm{fs}}''(\mathbf{r}=0)]=-q^2\gamma/5$. Finally, the resulting sum is converted to an integral (using $w\gg a$), yielding $\sum V_n^0=\sqrt{\pi}w/a$.

The second term can be shown to vanish as follows.
Since one can always write $\left.\partial^2_z u_{\mu}(\mathbf{r})\right|_{\mathbf{r}^{(0)}_n}=K_n u_{\mu}(\mathbf{r}^{(0)}_n)$ with some function of $n$ $K_n$, then from Eq. (\ref{26D}) we have $D''_{nm}=K_n D_{nm}$. Therefore, using $\sum_n\sum_m g_n K_n\mathrm{Re}[D_{nm}]g_m=0$ (can be shown from condition \ref{30a}), we find $\sum_n\sum_m g_n\mathrm{Re}[D''_{nm}]g_m=0$. Since $g_n\propto\sqrt{V_n^0}$, this means that the second term in (\ref{C3}) also vanishes.

For the last term, the fact that $V_n^0\propto g_n^2$ is paraxial implies $\Delta_{\mathbf{k}_{\bot}}\approx \Delta$ and the resulting sums simplify to $(1/N)\sum_{\mathbf{k}_{\bot}}\sum_{\mathbf{k}'_{\bot}}e^{i(\mathbf{k}_{\bot}-\mathbf{k}'_{\bot})\cdot\mathbf{r}_n^{\bot}}
\gamma_{\mathbf{k}_{\bot}\mathbf{k}'_{\bot}}=\gamma_{nn}$. Since $\gamma_{nn}$ represents the decay rate of a single atom to the non-confined modes, it does not vanish and is largely dominated by the decay in free space for which $\gamma_{nn}=\gamma$.

Using the above, we arrive at $\kappa_{sc}$ form Eq. (\ref{ksc}).

\subsection{Motion-induced decay: Estimation of $\kappa_{sc}$ from Eq. (\ref{112})}
The definition  of $\kappa_{sc}$ given in Eq. (\ref{112}) can be written as
\begin{eqnarray}
\kappa_{sc}&=&\kappa_1-\kappa_2,
\nonumber\\
\kappa_1&=&-2\mathrm{Re}\left[\sum_{\nu}C_{\nu\nu}\right], \quad \kappa_2=-2\mathrm{Re}[C_{00}].
\label{C16}
\end{eqnarray}
Starting with $\kappa_1$, we use $C_{\nu\nu}$ from Eq. (\ref{95}) and the fact that the collective mechanical mode profiles $V_n^0$ are real and orthonormal, finding
\begin{eqnarray}
\kappa_{1}&=&
2\sin^2(qz_0)\bar{g}\eta^2\sum_n V_n^0\frac{-\mathrm{Re}[D''_{nn}]}{q^2(\delta-\Delta)}
\nonumber\\
&+&\cos^2(qz_0)\bar{g}\eta^2\sum_nV_n^0
\frac{1}{N}
\nonumber\\
&&\times\mathrm{Re}\left[\sum_{\mathbf{k}_{\bot}}\sum_{\mathbf{k}'_{\bot}}e^{i(\mathbf{k}_{\bot}-\mathbf{k}'_{\bot})\cdot\mathbf{r}_n^{\bot}}
\frac{\gamma_{\mathbf{k}_{\bot}\mathbf{k}'_{\bot}}(\delta-\Delta)}{(\delta-\Delta_{\mathbf{k}_{\bot}})^2}\right].
\nonumber\\
\label{C18}
\end{eqnarray}
This is identical to the expression for $\kappa_{sc}$ in Eq. (\ref{C3}), considering that the second line in the latter (which is absent here) vanishes, as explained above. Therefore, we are left to show that $\kappa_2=0$, so that $\kappa_{sc}$ defined here [and in Eq. (\ref{112})] becomes identical to the result (\ref{ksc}) found above [starting from Eq. (\ref{C3})].

To show that $\kappa_2=0$ we use Eq. (\ref{95}) to write $\mathrm{Re}[C_{00}]$, finding that it is comprised of two terms which are proportional to the two quantities, respectively,
\begin{eqnarray}
\sum_{n,m}v_n\mathrm{Re}[D_{nm}]v_m, \quad \sum_{n,m}v_n\mathrm{Re}[D''_{nm}]v_m,
\label{qq}
\end{eqnarray}
where $v_n=(V_n^0)^{3/2}\propto g_n^3$. If $v_n$ was proportional to the cavity-mode profile $\sqrt{V_n^0}\propto g_n\propto u_{1,n}$ then we would be done, using the vanishing of the quantities $\sum_{n,m}g_n\mathrm{Re}[D_{nm}]g_m$ and $g_m\sum_{n,m}v_n\mathrm{Re}[D''_{nm}]g_m$, as explained above using the condition (\ref{30a}). However, recall that every mode whose transversely confined profile matches the cavity modes, should also satisfy condition (\ref{30a}). Since $v_n$, as a higher power of the Gaussian mode $g_n$, vastly overlaps with cavity-confined modes, it should approximately satisfy (\ref{30a}) as well. From here, we can show how the quantities in (\ref{qq}) are also vanishingly small, leading to $\kappa_2=0$.

\subsection{Second-order optomechanical coupling $g_2$}
For the evaluation of $g_2=-\mathrm{Im}[C_{00}]$, we use Eq. (\ref{95}) to find
\begin{eqnarray}
\mathrm{Im}[C_{00}]&=&\bar{g}\eta^2\left[\sin^2(qz_0)-\cos^2(qz_0)\right]\sum_n(V_n^0)^3
\nonumber\\
&+&\bar{g}\eta^2\frac{\sin^2(qz_0)}{q^2(\delta-\Delta)}\sum_{n,m}v_nv_m\mathrm{Im}[D''_{nm}],
\label{D5}
\end{eqnarray}
with $v_n=(V_n^0)^{3/2}$. For the first line, we convert the sum into an integral ($w\gg a$) finding $\sum_n (V_n^0)^3=\frac{4}{3}\frac{a}{\sqrt{\pi}w}$, whereas the second line is negligible as we explain below. This yields $g_2$ from Eq. (\ref{g2}).

To estimate the second line, we approximate the dispersive, near-field-dominated part of the Green's function by that of free space, $\mathrm{Im}[D''_{nm}]\approx \mathrm{Im}[\partial_z^2 D^{\mathrm{fs}}_{nm}]$. Using the representation of the free-space Green's function (\ref{Gfs}) in transverse 2D Fourier modes (see e.g. Ref. \cite{NH}, Eqs. 2.118 and 2.84),
\begin{eqnarray}
\left[\overline{\overline{G}}_{\mathrm{fs}}(\mathbf{r})\right]_{ij}=i\int\frac{d\mathbf{k}_{\bot}}{(2\pi)^2}e^{i\mathbf{k}_{\bot}\cdot\mathbf{r}}
\frac{e^{ik_z|z-z'|}}{2 k_z}\left(\delta_{ij}-\frac{k_i k_j}{q^2} \right)
\nonumber\\
\label{Gk}
\end{eqnarray}
with $k_z=\sqrt{q^2-|\mathbf{k}_{\bot}|^2}$ and where $k_i=\mathbf{k}\cdot \mathbf{e}_i$ is the projection of $\mathbf{k}=(\mathbf{k}_{\bot},k_z)$ along $i\in\{x,y,z\}$, we find
\begin{eqnarray}
&&\sum_{n,m}v_n v_m \mathrm{Im}[D''_{nm}]=-\frac{3}{4}\gamma\lambda N
\nonumber\\
&&\times\int \frac{d\mathbf{k}_{\bot}}{(2\pi)^2}v^2_{\mathbf{k}_{\bot}}
\mathrm{Im}\left[\sqrt{q^2-|\mathbf{k}_{\bot}|^2}\left(1-\frac{|\mathbf{k}_{\bot}\cdot\mathbf{e}_d|^2}{q^2}\right)\right].
\nonumber\\
\label{D7}
\end{eqnarray}
Here $v_{\mathbf{k}_{\bot}}=(1/\sqrt{N})\sum_ne^{-i\mathbf{k}_{\bot}\cdot \mathbf{r}^{\bot}_n}v_n$ is the Fourier transform of the paraxial profile $v_n$, which is only significant at small wavenumbers $|\mathbf{k}_{\bot}|\ll q$. On the other hand, the expression inside $\mathrm{Im}[...]$ is non-vanishing only if the argument inside the square root $\sqrt{q^2-|\mathbf{k}_{\bot}|^2}$ is negative, requiring $|\mathbf{k}_{\bot}|> q$, in contrast to the restriction  $|\mathbf{k}_{\bot}|\ll q$ imposed by $v_{\mathbf{k}_{\bot}}$. Therefore, $\sum_{n,m}v_n v_m \mathrm{Im}[D''_{nm}]$ in Eq. (\ref{D7}) is vanishingly small and so does the second line in (\ref{D5}).


\begin{thebibliography}{}
\bibitem{note2} Higher diffraction (Bragg) orders are suppressed for sufficiently small $a\lesssim \lambda$ \cite{coop}.
\bibitem{noteWG} For a recent example see Ref. \cite{ANA}, where waveguide QED with a 1D atom array was considered and cooperative effects mediated by non-guided modes were accounted for. This required considering a specific waveguide structure for which one could evaluate the non-guided part of the Green's function.
\bibitem{note1} This holds for every $\mathbf{k}_{\bot}$ ($|\mathbf{k}_{\bot}|<2\pi/\lambda$) if $a<\lambda/2$. For $\mathbf{k}_{\bot}\approx 0$ the less restrictive condition, $a<\lambda$, is sufficient.
\bibitem{note} Such an approximation should hold for experiment times shorter than $B_{\Delta}^{-1}$, where $B_{\Delta}$ is the frequency band spanned by the dispersion curve $\Delta_{\mathbf{k}_{\bot}}$ within the range of paraxial $\mathbf{k}_{\bot}$ values. In any case, $B_{\Delta}$ cannot be larger than the frequency span of the entire band, i.e. within the Brillouin zone $k_{x,y}\in \{-\pi/a,\pi/a\}$.
\end{thebibliography}
\end{document}